\documentclass{article}

\usepackage{amsmath,amsthm,amscd,amsxtra,amsfonts,amssymb,latexsym}

\begin{document}

\newcommand{\TeXButton}[2]{#2}
\newcommand{\stackunder}[2]{\underset{#1}{#2}}
\newcommand{\NEG}{\not}
\newcommand{\intmul}{\vdash}

\setcounter{secnumdepth}{2}
\setcounter{tocdepth}{1}

\title{Local moment maps and the splitting of classical multiplets}
\author{Hanno Hammer \\
Department of Applied Mathematics and Theoretical Physics,\\
Cambridge, Silver Street, UK}
\maketitle

\begin{abstract}
We generalize the concept of global moment maps to local moment maps
 whose different branches are labelled by the elements of the
 fundamental group of the underlying symplectic manifold. These
 branches can be smoothly glued together by employing
 fundamental-group-valued \u Cech cocycles on the phase space. In the
 course of this work we prove a couple of theorems on the liftability of
 group actions to symplectic covering spaces, and examine the possible
 extensions of the original group by the fundamental group of the
 quotient phase space. It it shown how the splitting of multiplets,
 this being a consequence of the multiply-connectedness of the
 quotient phase space, can be described by identification maps on a
 space of multiplets derived from a symplectic universal covering
 manifold. The states that are identified in this process are related
 by certain integrals over non-contractible loops in the quotient
 phase spacec.
\end{abstract}

\tableofcontents

\section{Introduction}

For a physical system whose Lagrangian is invariant under the action of the
isometry group $G$ of a background spacetime, the algebra of conserved
charges coincides with the Noether charge algebra of the system associated
with the isometry group. In recent years, it was observed by many authors
working in the field of high energy physics, that when classical or quantum
fields propagate in background spacetimes which are topologically
non-trivial then the associated algebra of conserved charges may reflect
this non-triviality by exhibiting extensions of the Noether charge algebra
which measure the topology of the background, provided the system is only
semi-invariant under the isometry group of the spacetime (see, e.g., \cite
{AzTown}). On the other hand, the algebra of conserved charges defines a
partition of the underlying phase space by distinguishing those subsets of
the phase space on which the values of all conserved charges involved are
constant (this is just the first step to a Marsden-Weinstein reduction).
These subsets then could be called ''elements of classical $G$-multiplets'',
a multiplet being defined as a $G$-orbit of such subsets. On every such
multiplet, the symmetry group $G$ acts transitively, thus exhibiting a
property which is the classical analogue of an irreducible representation in
the quantum theory \cite{Woodhouse}. Now, if dynamical systems are described
in the framework of symplectic formalism, the quantity determining this
phase space partition is what we call a global moment map on the phase
space. However, this global moment map need not exist in the case that the
phase space is not simply connected. This observation was the starting point
for the investigations in this work. For systems with finite degrees of
freedom, we generalize the concept of a global moment map to multiply
connected phase spaces and general symplectic manifolds. We show that the
appropriate generalization involves a locally defined multi-valued moment
map, whose different branches are labelled by the fundamental group of the
phase space. Furthermore, it is shown how the different local branches can
be smoothly glued together by a glueing condition, which is expressed by
certain \u Cech cocycles on the underlying symplectic manifold. These
constructions are intimately related with the existence of a universal
symplectic covering manifold of the original phase space. On the covering,
global moment maps defining $G$-multiplets always exist. At the end of this
work we show how these multiplets on the covering space can be related to $G$%
-multiplets on the original symplectic manifold by an identification map
which derives from the covering projection.

In order to formulate these ideas rigourosly we first had to examine the
question of liftability of symplectic group actions on a symplectic manifold
to a covering space. In the course of this, we proved a series of theorems
investigating the existence and uniqueness of such lifts, and to which
extent the lifted action preserves the group structure of the original
symmetry group $G$.

The plan of the paper is as follows: In section \ref{BasicsAboutCoverings}
we collect basic statements about covering spaces on which the rest of this
work relies. In section \ref{CoveringsAndDiffForms} and \ref{MultiValFunc}
we examine multi-valued potential functions for closed but not exact
differential forms on a multiply connected manifold. These considerations
will be needed later on when examining local moment maps. In section \ref
{NotationSymplecticManifolds}--\ref{CotangentBundles} we collect notation
conventions and basic facts about symplectic manifolds, Hamiltonian vector
fields, and cotangent bundles. In section \ref{CoveringCotBundle} we study
how covering projections can be extended to local symplectomorphisms of
coverings of cotangent bundles. Sections \ref{LiftOfGroupActions} and \ref
{PreservationOfTheGroupLaw} examine the conditions under which an action of
a Lie group $G$ on a manifold can be lifted to an action on a covering
manifold, and when such a lift preserves the group law of $G$. Section \ref
{GSpacesCoveringMaps} examines the relation between group actions on
covering spaces and equivariance of the covering map. In \ref
{SymplecticGActionsMoMaps} we introduce our notation for symplectic $G$%
-actions on a symplectic manifold. Section \ref{GlobalMomentMaps}
recapitulates the notion of global moment maps as they are usually defined,
while this concept is generalized to local moment maps in the subsequent
section \ref{LocalMomentMaps}. Equivariance of global and local moment maps
is discussed in section \ref{EquivarianceOfMomentMaps}. Section \ref
{NonSimplyConnected} discusses the relation of local moment maps to covering
spaces which are themselves multiply connected. In \ref{GStateSpaces} we
introduce the concept of $G$-states, while the subsequent section \ref
{SplittingOfMultiplets} shows how a splitting of $G$-states on multiply
connected symplectic manifold arises.

\section{Basic facts about coverings \label{BasicsAboutCoverings}}

In this section we quote the main results on coverings and lifting theorems
on which the rest of this work is built; it is mainly based on \cite{Fulton},%
\cite{Jaehnich},\cite{ONeill},\cite{Wolf}. In this work we are interested in
covering spaces that are manifolds. Consequently, some of the definitions
and quotations of theorems to follow are not presented in their full
generality, as appropriate for general topological spaces, but rather we
give working definitions and formulations pertaining to manifolds and the
fact that these are special topological spaces (locally homeomorphic to $%
\TeXButton{R}{\mathbb{R}}^n$).

--- A covering of manifolds is a triple $\left( p,X,Y\right) $, where $%
p:X\rightarrow Y$ is a smooth surjective map of smooth manifolds $X,Y$,
where $Y$ is connected, such that for every $y\in Y$ there exists an open
neighbourhood $V\subset Y$ of $y$ for which $p^{-1}\left( V\right) $ is a
disjoint union of open sets $U$ in $X$, on each of which the restriction $%
\left. p\right| U:U\rightarrow Y$ is a diffeomorphism. Every open $V\subset
Y $ for which this is true is called admissible (with respect to $p$). This
means that, for each $y\in Y$, the inverse image $p^{-1}\left( y\right)
\subset X$, called the fibre over{\it \ }$y$, is discrete. Since $Y$ is
connected, all fibres have the same cardinality. $p$ is called projection or
covering map. A diffeomorphism $\phi :X\rightarrow X$ such that $p\circ \phi
=p$ is called a deck{\it \ }transformation of the covering. The set ${\cal D}
$ of all deck transformations of the covering is a group under composition
of maps. Since every $\phi \in {\cal D}$ permutes the elements in the fibres 
$p^{-1}\left( y\right) $, the group ${\cal D}$ of all $\phi $ is discrete.
If $X$ is connected, deck transformations are uniquely determined by their
value at a given point $x\in X$.

--- Let $x\in p^{-1}\left( y\right) $, and let $\pi _1\left( X,x\right) $, $%
\pi _1\left( Y,y\right) $ denote the fundamental groups of $X$ and $Y$ based
at $x$ and $y$, respectively. The projection $p$ induces a homomorphism $%
p_{\#}:\pi _1\left( X,x\right) \rightarrow \pi _1\left( Y,y\right) $ of
fundamental groups, such that $p_{\#}\pi _1\left( X,x^{\prime }\right) $
ranges through the set of all conjugates of $p_{\#}\pi _1\left( X,x\right) $
in $\pi _1\left( Y,y\right) $, as $x^{\prime }$ ranges through the elements
in the fibre $p^{-1}\left( y\right) $. A covering $p:X\rightarrow Y$ is
called normal if $p_{\#}\pi _1\left( X,x^{\prime }\right) $ is normal in $%
\pi _1\left( Y,y\right) $ for some (hence any) $x^{\prime }\in p^{-1}\left(
y\right) $. One can show that a covering is normal if and only if the deck
transformation group ${\cal D}$ acts transitively on the fibres, i.e. for
all $x,x^{\prime }\in X$ with $p\left( x\right) =p\left( x^{\prime }\right) $
there exists a unique deck transformation $\phi \in {\cal D}$ with $%
x^{\prime }=\phi \left( x\right) $. This is certainly true when $X$ is
simply connected; in this case, the deck transformation group ${\cal D}$ is
isomorphic to the fundamental group $\pi _1\left( Y,y\right) $.

--- Let $\Gamma $ be a group of diffeomorphisms acting on the manifold $X$,
and let $\Gamma x$ denote the orbit of $x$ under $\Gamma $. The set of all
orbits is denoted by $X/\Gamma $, and is called an orbit space. The natural
projection $pr:X\rightarrow X/\Gamma $ sends each $x\in X$ to its orbit, $%
pr\left( x\right) =\Gamma x$. The topology on $X/\Gamma $ is the quotient
topology, for which $p$ is continuous, and an open map. $\Gamma $ acts
properly discontinuously on $X$ if every $x\in X$ has a neighbourhood $U$
such that the set $\left\{ \gamma \in \Gamma \mid \gamma U\cap U\neq
\emptyset \right\} $ is finite. $\Gamma $ acts freely if no $\gamma \neq e$
has a fixed point in $x$. $\Gamma $ acts properly discontinuously and freely
if each $x\in X$ has a neighbourhood $U$ such that $e\neq \gamma \in \Gamma $
implies $\gamma U\cap U=\emptyset $.

--- Given a covering $p:X\rightarrow Y$ and a map $f:V\rightarrow Y$ defined
on some manifold $V$, a map $\tilde f:V\rightarrow X$ is called a lift of $f$%
\ through $p$ if $p\circ \tilde f=f$. Two maps $f,g:V\rightarrow Y$ are
called homotopic if there exists a continuous map $G:I\times V\rightarrow Y$%
, $\left( t,v\right) \mapsto G_t\left( v\right) $, such that $G_0=f$ and $%
G_1=g$. $G$ is called a homotopy of $f$\ and $g$.

\TeXButton{Abst}{\vspace{0.8ex}}--- We now quote without proof a couple of
theorems from \cite{Wolf} which we will make us of frequently. These
theorems are actually proven in every textbook on Algebraic Topology.

{\bf Theorem:} \label{TheoWolf1}\quad If $\Gamma $ acts properly
discontinuously and freely on the connected manifold $X$, the natural
projection $pr:X\rightarrow X/\Gamma $ onto the orbit space is a covering
map, and the covering is normal. Furthermore, the deck transformation group $%
{\cal D}$ of this covering is $\Gamma $.

\TeXButton{Abst}{\vspace{0.8ex}}As a converse, we have

{\bf Theorem:} \label{TheoWolf2}\quad If $p:X\rightarrow Y$ is a covering
and ${\cal D}$ is the group of deck transformations, then ${\cal D}$ acts
properly discontinuously and freely on $X$. --- If $X$ is a simply connected
manifold, every covering $p:X\rightarrow Y$ is a natural projection $%
pr:X\rightarrow X/\Gamma $ for some discrete group $\Gamma $ of
diffeomorphisms acting properly discontinuously and freely on $X$.

\TeXButton{Abst}{\vspace{0.8ex}}As a consequence, we have

{\bf Corollary:} \label{CoroWolf3}\quad Let $Y$ be a connected manifold.
Then $Y$ is diffeomorphic to an orbit space $X/\Gamma $, where $X$ is a
simply connected covering manifold $X$, and $\Gamma $ is a group of
diffeomorphisms acting properly discontinuously and freely on $X$. In this
case, $\Gamma ={\cal D}$ coincides with the deck transformation group of the
covering.

\TeXButton{Abst}{\vspace{0.8ex}}--- Existence and uniqueness of lifts are
determined by the following lifting theorems:

{\bf Theorem (''Unique Lifting Theorem''):} \label{UniqueLiftingTheorem}%
\quad Let $p:X\rightarrow Y$ be a covering of manifolds. If $V$ is a
connected manifold, $f:V\rightarrow Y$ is continuous, and $g_1$ and $g_2$
are lifts of $f$ through $p$ that coincide in one point, $g_1\left( v\right)
=g_2\left( v\right) $, then $g_1=g_2$.

\TeXButton{Abst}{\vspace{0.8ex}}{\bf Theorem (''Covering Homotopy Theorem''):%
} \label{CoveringHomotopyTheorem}\quad Let $p:X\rightarrow Y$ be a covering.
Let $f,g:V\rightarrow Y$ be continuous, let $\tilde f:V\rightarrow X$ be a
lift of $f$, and let $G$ be a homotopy of $f$ and $g$. Then there is a
unique pair $\left( \tilde G,\tilde g\right) $, where $\tilde g$ is a lift
of $g$, $\tilde G$ is a lift of $\tilde G$, and $\tilde G$ is a homotopy of $%
\tilde f$ and $\tilde g$.

\TeXButton{Abst}{\vspace{0.8ex}}{\bf Theorem (''Lifting Map Theorem''):} 
\label{LiftingMapTheorem}\quad Let $p:X\rightarrow Y$ be a covering, let $V$
be a connected manifold, let $f:V\rightarrow Y$ be continuous. Let $x\in X$, 
$y\in Y$, $v\in V$ such that $p\left( x\right) =y=f\left( v\right) $. Then $%
f $ has a lift $\tilde f:V\rightarrow X$ with $\tilde f\left( v\right) =x$
if and only if $f_{\#}\pi _1\left( V,v\right) \subset p_{\#}\pi _1\left(
X,x\right) $, where $f_{\#}$ denotes the homomorphism of fundamental groups
induced by $f$.

\section{Covering spaces and differential forms \label{CoveringsAndDiffForms}
}

Let $p:X\rightarrow Y$ be a covering of smooth manifolds. In this case the
covering projection $p$ is a local diffeomorphism. In this section we
examine the relation between differential forms on $X$ and $Y$ which are
related by pull-back through the covering map $p$, and where $X$ is simply
connected. The results will be needed to formulate the concept of local
moment maps on non-simply connected symplectic manifolds in section \ref
{LocalMomentMaps}.

\subsection{Differential forms on $X$ and $Y$}

First consider a smooth $q$-form $\omega $ on $Y$. The pull-back $\Omega
\equiv p^{*}\omega $ is a $q$-form on $X$, and $\Omega $ is invariant under
the deck transformation group of the covering: For, let $\gamma \in {\cal D}$%
, then $\gamma ^{*}\Omega =\left( p\circ \gamma \right) ^{*}\omega =\Omega $%
. Conversely, let $\Omega $ be a $q$-form on $X$. Then, locally, $\Omega $
can be pulled back to $Y$ to give a multi-valued $q$-form on $Y$, since the
covering map $p$ is a local diffeomorphism: To see this, let $V$ be an
admissible neighbourhood of a point $y\in Y$, such that the inverse image $%
p^{-1}\left( V\right) $ is a disjoint union of neighbourhoods $U_i\,$in $X$.
On each $U_i$, the restriction $\left. p\right| U_i:U_i\rightarrow Y$ is a
diffeomorphism, so that we can pull back $\left. \Omega \right| U_i\mapsto
\left( \left. p\right| U_i\right) ^{-1*}\Omega $ on $V$. This gives a
multi-valued $q$-form on $V$, each branch being labelled by some $i$. We ask
under which condition all branches coincide. If this happens to be, we have $%
\left( \left. p\right| U_i\right) ^{-1*}\Omega =\left( \left. p\right|
U_j\right) ^{-1*}\Omega $ for all $i,j$ labelling different neighbourhoods $%
U_i$, $U_j$; but this means that 
\begin{equation}
\label{pp6t1form1}\left[ \left( \left. p\right| U_i\right) ^{-1}\circ \left(
\left. p\right| U_j\right) \right] ^{*}\left( \left. \Omega \right|
U_i\right) =\left( \left. \Omega \right| U_j\right) \quad . 
\end{equation}
For a general covering we can proceed no further, since there need not exist
a deck transformation mapping the neighbourhoods $U_i$ and $U_j$ into each
other. Such a deck transformation exists, however, if $X$ is simply
connected. If $x_i\in U_i$ and $x_j\in U_j$ such that $p\left( x_j\right)
=p\left( x_i\right) =y$, then there is a unique deck transformation $\gamma $
with $x_j=\gamma \left( x_i\right) $. On the other hand, the map on the LHS
of (\ref{pp6t1form1}) satisfies 
\begin{equation}
\label{pp6t1form2}\left( \left. p\right| U_i\right) \circ \left[ \left(
\left. p\right| U_i\right) ^{-1}\circ \left( \left. p\right| U_j\right)
\right] =\left( \left. p\right| U_j\right) 
\end{equation}
and maps $x_j$ to $x_i$; hence it coincides with the restriction $\left(
\gamma \mid U_j,U_i\right) $. This argument can be performed for any
neighbourhood $U$ of some point $x\in X$; in turn, this implies that $\gamma
^{*}\Omega =\Omega $. The fact that ${\cal D}$ now acts transitively on the
fibres of the covering means that this relation must hold for all deck
transformations $\gamma \in {\cal D}$.

As a consequence, if a $q$-form $\Omega $ on $X$ is invariant under ${\cal D}
$, then there is a uniquely determined $q$-form $\omega $ on $Y$ such that $%
\Omega =p^{*}\omega $; for, $\omega $ is defined by mapping any vector $W\in
T_yY$ to $\left( p_{loc}\right) _{*}W$, where $p_{loc}$ denotes the
restriction of $p$ to one of the connected components of the inverse image
of an admissible neighbourhood of $y$ in $Y$, and subsequently performing
the pairing $\left\langle \Omega ,\left( p_{loc}\right) _{*}W\right\rangle $%
. Because of the ${\cal D}$-invariance of $\Omega $, it does not matter
which connected component we choose, and hence this construction is
well-defined.

Altogether, we have shown

\subsection{Proposition \label{DeckInvariant}}

Let $p:X\rightarrow Y$ be a covering of connected manifolds, where $X$ is
simply connected. Then a smooth $q$-form $\Omega $ on $X$ is the pull-back
of a $q$-form $\omega $ on $Y$, $\Omega =p^{*}\omega $, if and only if $%
\gamma ^{*}\Omega =\Omega $ for all $\gamma \in {\cal D}$.

\TeXButton{Abst}{\vspace{0.8ex}}We now discuss the closure and exactness of
forms on $X$ and $Y$ that are related by the covering map according to $%
\Omega =p^{*}\omega $. Since closure of a differential form is a local
property, $\Omega $ is closed if and only if $\omega $ is closed. For if $%
d\omega =0$, then $dp^{*}\omega =p^{*}d\omega =0$; and conversely, if $V$ is
an admissible neighbourhood in $Y$, $d\left( \left. \omega \right| V\right)
=d\left[ \left( \left. p\right| U_i\right) ^{-1*}\Omega \right] =0$, if $%
d\Omega =0$; here $U_i$ is any neighbourhood in $X$ that projects down to $V$%
. This result is actually independent of whether $X$ is simply connected or
not, and makes use only of the existence of a form $\Omega $ such that $%
\Omega =p^{*}\omega $.

Now we examine exactness. Trivially, if the form $\omega $ on $Y$ is exact,
then $\Omega $ is exact, since then $\Omega =p^{*}d\alpha =d\left(
p^{*}\alpha \right) $. The converse is not true in general, however. For
assume that $\Omega =d\eta $ for a $\left( q-1\right) $-form $\eta $ on $X$.
Assuming that $\Omega =p^{*}\omega $, proposition \ref{DeckInvariant} says
that $d\left( \gamma ^{*}\eta -\eta \right) =0$. Thus $\gamma ^{*}\eta -\eta 
$ is closed, and since $X$ is simply connected, it is also exact. This means
that there exists a $\left( q-2\right) $-form $\chi \left( \gamma \right) $
on $X$ such that 
\begin{equation}
\label{pp6t1form3}\gamma ^{*}\eta =\eta +d\chi \left( \gamma \right) \quad . 
\end{equation}
Now unless $d\chi =0$, we see from proposition \ref{DeckInvariant} that $%
\eta $ can{\bf not} be the pull-back of a $\left( q-1\right) $-form on $Y$
under $p$, as this requires $\eta $ to be ${\cal D}$-invariant. Thus,
although $\Omega $ is exact, $\omega $ need not be exact; it is closed,
however, and hence defines an element $\left[ \omega \right] \in
H_{deRham}^q\left( Y\right) $.

The content of the last two paragraphs can be cast into a convenient form by
introducing ${\cal D}${\it -invariant cohomology classes }of forms on $X$:

\subsection{Definition}

Let $\Lambda _{{\cal D}I}^q\left( X\right) $ denote the subspace of all $%
{\cal D}$-invariant $q$-forms on $X$, i.e. $\gamma ^{*}\Omega =\Omega $ for
all $\gamma \in {\cal D}$. Let $Z_{{\cal D}I}^q\left( X\right) $ denote the
class of all elements $\Omega $ in $\Lambda _{{\cal D}I}^q\left( X\right) $
which are closed under $d$, $d\Omega =0$. Let $B_{{\cal D}I}^q\left(
X\right) $ denote the class of forms $\Omega $ in $\Lambda _{{\cal D}%
I}^q\left( X\right) $ which are exact under $d$, i.e. there exists a $\eta
\in \Lambda _{{\cal DI}}^{q-1}\left( X\right) $ such that $\Omega =d\eta $.
Now define the $q$-th ${\cal D}${\it -invariant cohomology group} $H_{{\cal D%
}I}^q\left( X\right) $ on $X$ as the quotient 
\begin{equation}
\label{pp6t1form4}H_{{\cal D}I}^q\left( X\right) \equiv Z_{{\cal D}%
I}^q\left( X\right) /B_{{\cal D}I}^q\left( X\right) \quad . 
\end{equation}
Formula (\ref{pp6t1form3}) shows that $B_{{\cal D}I}^q\left( X\right) 
\stackunder{\neq }{\subset }Z_{{\cal D}I}^q\left( X\right) $ in general, and
so $\dim H_{{\cal D}I}^q\left( X\right) $ can be non-vanishing although $X$
has trivial de Rham cohomology groups. This is expressed in the next
proposition, which is a consequence of proposition \ref{DeckInvariant} and
the discussion in the last paragraph:

\subsection{Proposition \label{DInvariantPullBack}}

Assume that $X$ is simply connected. Then the pull-back $p^{*}\Lambda
^q\left( Y\right) $ of the space of $q$-forms $\Lambda ^q\left( Y\right) $
on $Y$ by the covering map $p$ coincides with the subspace of all ${\cal D}$%
-invariant $q$-forms $\Lambda _{{\cal D}I}^q\left( X\right) $ on $X$, and $%
p^{*}$ is a group isomorphism onto $\Lambda _{{\cal D}I}^q\left( X\right) $.
Furthermore, $Z_{{\cal D}I}^q\left( X\right) =p^{*}Z^q\left( Y\right) $ and $%
B_{{\cal D}I}^q\left( X\right) =p^{*}B^q\left( Y\right) $, and therefore 
\begin{equation}
\label{pp6t1form5}\dim H_{{\cal D}I}^q\left( X\right) =\dim \frac{%
p^{*}Z^q\left( Y\right) }{p^{*}B^q\left( Y\right) }=\dim \frac{Z^q\left(
Y\right) }{B^q\left( Y\right) }=\dim H^q\left( Y\right) \;>\;0 
\end{equation}
in general.

\TeXButton{Abst}{\vspace{0.8ex}}There is a cohomological description of
formula (\ref{pp6t1form3}) in terms of special $\Lambda ^{*}\left( X\right) $%
-valued cochains, where $\Lambda ^{*}\left( X\right) $ denotes the ring of
differential forms on $X$; the associated cohomology is defined and
described in the appendix, section \ref{FormValuedCohomologyOnD}. For the
work pursued here the general case has no immediate application, but the
case when $X$ is simply connected and the forms involved are $1$-forms is
important. To start, let $\alpha $ be a closed $1$-form on $Y$; then $%
p^{*}\alpha $ is closed on $X$, hence exact, since $X$ is simply connected.
Thus there exists a smooth function $F:X\rightarrow \TeXButton{R}{\mathbb{R}}
$ with $dF=p^{*}\alpha $. The discussion in proposition \ref{DeckInvariant}
has shown that $dF$ is ${\cal D}$-invariant; therefore, $d\left( \gamma
^{*}F-F\right) =0$, or 
\begin{equation}
\label{pp6t1form6}F\circ \gamma -F\equiv c\left( \gamma \right) \in 
\TeXButton{R}{\mathbb{R}} 
\end{equation}
is a real constant on $X$, depending only on $\gamma $. In particular, $%
c\left( \gamma \right) \circ \gamma ^{\prime }=c\left( \gamma \right) $, and
it follows that 
\begin{equation}
\label{pp6t1form7}c\left( \gamma \gamma ^{\prime }\right) =F\circ \left(
\gamma \gamma ^{\prime }\right) -F=\left[ F+c\left( \gamma \right) \right]
\circ \gamma ^{\prime }-F=c\left( \gamma \right) +c\left( \gamma ^{\prime
}\right) \quad ; 
\end{equation}
hence $c:{\cal D}\rightarrow \TeXButton{R}{\mathbb{R}}$, $\gamma \mapsto
c\left( \gamma \right) $ is a real $1$-dimensional representation of ${\cal D%
}$. We have proven:

\subsection{Proposition \label{Sheets}}

Let $p:X\rightarrow Y$ be a covering of smooth manifolds, where $X$ is
simply connected. Let $\alpha $ be a closed $1$-form on $Y$. Then the
pull-back $p^{*}\alpha $ is an exact $\gamma $-invariant $1$-form on $X$,
with $p^{*}\alpha =dF$, where $F\in {\cal F}\left( X\right) $ is a smooth
function on $X$. Under ${\cal D}$-transformations, $F$ is invariant up to a
real $1$-dimensional ${\cal D}$-representation $c:{\cal D}\rightarrow 
\TeXButton{R}{\mathbb{R}}$, i.e. 
\begin{equation}
\label{pp6t1form8}F\circ \gamma =F+c\left( \gamma \right) \quad . 
\end{equation}

\TeXButton{Abst}{\vspace{0.8ex}}Now we see how the function $F$ gives rise
to multi-valued locally defined functions $f_\gamma $ on $Y$, which
represent local potentials, i.e. $0$-forms, for the closed $1$-form $\alpha $%
: Given an admissible open neighbourhood $V\subset Y$, choose a connected
component $U\subset X$ of $p^{-1}\left( V\right) $; then $\left. p\right| U$
is a diffeomorphism onto $V$, and every other connected component in $%
p^{-1}\left( V\right) $ is obtained as the image of $U$ under a deck
transformation $\gamma $. Since the sets $\gamma U$, $\gamma \in {\cal D}$,
are disjoint, this determines a collection $\left( f_\gamma \right) $ of
local potentials for $\alpha $ on $V$, each $f_\gamma $ being defined as 
\begin{equation}
\label{pp6t1pot1}f_\gamma =F\circ \left( \left. p\right| \gamma U\right)
^{-1}\quad . 
\end{equation}
By construction, we have $df_\gamma =\alpha $ for every $\gamma \in {\cal D}$%
. Furthermore, since $\left( \left. p\right| \gamma U\right) \circ \gamma
=\left. p\right| U$, it follows that 
\begin{equation}
\label{pp6t1pot2}f_\gamma =F\circ \left[ \gamma \circ \left( \left. p\right|
U\right) ^{-1}\right] =\left[ F\circ \gamma \right] \circ \left( \left.
p\right| U\right) ^{-1}=\left[ F+c\left( \gamma \right) \right] \circ \left(
\left. p\right| U\right) ^{-1}=f_e+c\left( \gamma \right) \quad , 
\end{equation}
where we have used (\ref{pp6t1form8}).

\TeXButton{Abst}{\vspace{0.8ex}}We now prove the important result, that $%
c\left( \gamma \right) $ can be expressed as an integral of $\alpha $ over
certain $1$-cycles or loops in $Y$: To this end we recall that the deck
transformation group ${\cal D}$ acts properly discontinuously and freely on
the simply connected covering manifold $X$, and that $Y$ is the orbit space $%
X/{\cal D}$. Furthermore, if $x,y$ are base points of $X,Y$, with $p\left(
x\right) =y$, then the fundamental group $\pi _1\left( Y,y\right) $ of $Y$
at $y$ is isomorphic to ${\cal D}$. This isomorphism is defined as follows:
If $\gamma \in {\cal D}$, let $\lambda $ be an arbitrary path in $X$
connecting the base point $x$ with its image $\gamma \left( x\right) $; then 
$\lambda $ projects into a loop $p\circ \lambda $ at $y$, whose associated
homotopy class $\left[ p\circ \lambda \right] $ represents $\gamma \in \pi
_1\left( Y,y\right) $. Any other choice $\lambda ^{\prime }$ of path is
homotopic to $\lambda $ due to $X$ being simply connected, hence the loops $%
\left[ p\circ \lambda \right] $ and $\left[ p\circ \lambda ^{\prime }\right] 
$ both represent the same homotopy class. Conversely, given a loop $l$ at $y$
representing $\gamma $, there exists a unique lift $\tilde l$ of $l$ through 
$p$ with initial point $x\in p^{-1}\left( y\right) $ (as follows from
theorem \ref{UniqueLiftingTheorem}); this means that $p\circ \tilde l=l$,
and $\tilde l\left( 1\right) =\gamma \left( x\right) $. Now we can prove

\subsection{Theorem \label{Cozykel}}

Let $y\in Y$ be the base point of $Y$, let $l$ be a loop at $y$ with $\left[
l\right] =\gamma \in \pi _1\left( Y,y\right) \simeq {\cal D}$. Then 
\begin{equation}
\label{pp6t1fo9}c\left( \gamma \right) =\int\limits_l\alpha \quad . 
\end{equation}
The integral depends only on the homotopy class $\left[ l\right] $ of $l$.

\TeXButton{Beweis}{\raisebox{-1ex}{\it Proof :}
\vspace{1ex}}

The lift $\tilde l$ of $l$ to the base point $x$ of $X$ satisfies $\tilde
l\left( 1\right) =\gamma \left( x\right) $. Since $p\circ \tilde l=l$, we
have%
$$
\int\limits_l\alpha =\int\limits_{p\circ \tilde l}\alpha
=\int\limits_{\tilde l}p^{*}\alpha \quad , 
$$
but $p^{*}\alpha =dF$, and since $\tilde l$ connects $x$ and $\gamma \left(
x\right) $, the last integral in the above equation is%
$$
\int\limits_{\tilde l}p^{*}\alpha =\int\limits_x^{\gamma \left( x\right)
}dF=F\circ \gamma \left( x\right) -F\left( x\right) =\left[ F\circ \gamma
-F\right] \left( x\right) =c\left( \gamma \right) \quad , 
$$
according to the definition (\ref{pp6t1form8}) of $c\left( \gamma \right) $.
This proves (\ref{pp6t1fo9}). Any other loop $l^{\prime }$ homotopic to $l$
lifts to a path $\tilde l^{\prime }$ homotopic to $\tilde l$; hence the
difference between the associated integrals is an integral of $dF$ over a
loop, which must vanish, as $dF$ is exact. \TeXButton{BWE}
{\hfill
\vspace{2ex}
$\blacksquare$}

\subsection{Corollary}

Any element $\gamma $ of ${\cal D}$ that has torsion lies in the kernel of $%
c $; in other words, if there is a $k\in \TeXButton{N}{\mathbb{N}}$ with $%
\gamma ^k=e$ then $c\left( \gamma \right) =0$.

\TeXButton{Beweis}{\raisebox{-1ex}{\it Proof :}
\vspace{1ex}}

Insert $e=\gamma ^k$ into $c\left( e\right) =0$, which gives $0=c\left(
\gamma ^k\right) =k\cdot c\left( \gamma \right) $ due to (\ref{pp6t1form7}).
Since $k>0$, $c\left( \gamma \right) =0$. \TeXButton{BWE}
{\hfill
\vspace{2ex}
$\blacksquare$}

\section{\u Cech cohomology and multi-valued functions \label{MultiValFunc}}

We now want to make more precise the notion of multi-valued functions that
serve as local potentials for closed $1$-forms on the non-simply connected
manifold $Y$. We want to express the local potentials, as they were defined
in formula (\ref{pp6t1pot1}), and their mutual relations, in terms of
locally defined quantities and glueing conditions without explicit reference
to a specific covering manifold. It is clear that we must make up for the
information that is lost by discarding the covering space from consideration
by some additional structure on the manifold $Y$. It turns out that the
necessary ingredients are

\begin{description}
\item[1.)]  \quad a countable simply connected path-connected open cover $%
{\cal V=}\left\{ V_a\subset Y\mid a\in A\right\} $ of $Y$, i.e. a collection
of countably many open sets $V_a\subset Y$ whose union gives $Y$, and such
that every loop in $V_a$ is homotopic in $Y$ to a constant loop, and all $%
V_a $ are path-connected.
\end{description}

Since $Y$ is a manifold, a cover of the type just described always exists.
We note that, as every element $V_a\in {\cal V}$ is simply connected (in $Y$%
), it is automatically admissible with respect to the covering map $%
p:X\rightarrow Y$; for, assume it were not admissible; then the inverse
image $p^{-1}\left( V_a\right) $ contained a connected, hence
path-connected, component $U$ on which the restriction of $p$ is not
injective. In particular, there are points $x,x^{\prime }\in U$, $x\neq
x^{\prime }$, but $p\left( x\right) =p\left( x^{\prime }\right) $. Choose a
path $\lambda $ connecting $x$ and $x^{\prime }$ in $U$; then this projects
into a loop in $V_a$ at $p\left( x\right) $, which is non-contractible in $Y$%
. This contradicts the assumption of $V_a$ being simply connected. --- The
second ingredient is

\begin{description}
\item[2.)]  \quad a ${\cal D}$-valued $1$-\u Cech-cocycle $\left( g_{ab}\in 
{\cal D}\right) $, $a,b\in A$, on ${\cal V}$, satisfying a certain condition
which expresses that the class of ${\cal D}$-isomorphic covering spaces to
which it refers is simply connected (elements of \u Cech cohomology and its
relation to covering spaces are explained in the appendix, sections \ref
{CechCohomology}, \ref{DCoverings}, \ref{CechAndGlueing}).
\end{description}

We now explain what this condition means. Let $p:X\rightarrow Y$ be a
universal covering manifold of $Y$ with base points $x\in X$ and $y\in Y$
such that $y=p\left( x\right) $. The deck transformation group ${\cal D}$ of
such a covering is isomorphic to the fundamental group $\pi _1\left(
Y,y\right) $, the isomorphism being defined as in the discussion preceding
theorem \ref{Cozykel}: If $\left[ \lambda \right] $ is any loop class in $%
\pi _1\left( Y,y\right) $, let $\tilde \lambda $ denote the unique lift of $%
\lambda $ to $x$; then there exists a unique deck transformation $\gamma $
such that $\gamma x=\tilde \lambda \left( 1\right) $. This deck
transformation is the image of $\left[ \lambda \right] $ under the
above-mentioned isomorphism. The manifold $Y$ can be considered as the orbit
space $X/{\cal D}$ with the quotient topology. We now identify ${\cal D}$
with the fundamental group $\pi _1\left( Y,y\right) $, so that $Y$ can be
regarded as the orbit space $X/\pi _1\left( Y,y\right) $, $\pi _1$ acting
via deck transformations $\gamma $ on $X$, and $p:X\rightarrow X/{\cal D}=Y$
is a ${\cal D}$-covering.

Next, we note that, although a simply connected covering manifold $X$ of $Y$
is uniquely defined up to ${\cal D}$-isomorphisms, in general there are also 
${\cal D}$-coverings $q:Z\rightarrow Y=Z/{\cal D}$ of $Y$ which are {\bf not}
connected. This means that $Y$ can be expressed as an orbit space of
different, not necessarily connected, manifolds $Z$ (with base point $z$),
under an action of ${\cal D}$. These manifolds can be assembled into
equivalence classes, equivalence being expressed by ${\cal D}$-isomorphism
(see appendix, section \ref{DCoverings}), and a universal covering manifold $%
X$ determines just one class in this collection. As explained in the
appendix, section \ref{CechAndGlueing}, there is a bijection between these
equivalence classes and the classes of cohomologous $1$-\u Cech-cocycles on $%
{\cal V}$, in other words, the elements of $H^1\left( {\cal V;D}\right) $;
hence the bijection 
\begin{equation}
\label{pp6t1multValFunc10}\left\{ {\cal D}\text{-coverings with base point}%
\right\} /\text{isomorphism\ }\leftrightarrow \;H^1\left( {\cal V;D}\right)
\quad . 
\end{equation}
Furthermore, a result in the theory of covering spaces (see, e.g., \cite
{Fulton}) states that there is a bijection between $\left\{ {\cal D}\text{%
-coverings with base point}\right\} /$isomorphism and the set $Hom\left( 
{\cal D,D}\right) $ of homomorphisms from ${\cal D}=\pi _1\left( Y,y\right)
\rightarrow {\cal D}$. Hence we also have a bijection 
\begin{equation}
\label{pp6t1multValFunc11}H^1\left( {\cal V;D}\right) \;\leftrightarrow
\;Hom\left( {\cal D,D}\right) \quad . 
\end{equation}
We explain (LHS$\rightarrow $RHS) of this bijection. Let the \u Cech cocycle 
$\left( g_{ab}\right) $ be given. We first show that a ${\cal D}$-covering $%
Z $ of $Y$ exists such that the \u Cech cocycle determined by a collection $%
\left( i_a\right) $ of trivializations is the given one. To this end,
consider the topological sum of all ${\cal D}\times V_a$ (i.e. the
underlying set is a disjoint union) and define the relation $\left(
d,y\right) \sim \left( d^{\prime },y^{\prime }\right) $ for elements $\left(
d,y\right) \in {\cal D}\times V_a$, $\left( d^{\prime },y^{\prime }\right)
\in {\cal D}\times V_b$ to be true if and only if $\left( d^{\prime
},y^{\prime }\right) =\left( d\cdot g_{ab},y\right) $; then properties
(Trans1-Trans3) in the appendix, section \ref{CechAndGlueing}, guarantee
that $"\sim "$ is an equivalence relation. Now define $Z$ to be the quotient 
$Z\equiv \sqcup _{a\in A}{\cal D}\times V_a/\sim $, endowed with the final
topology (quotient topology). It is easy to see that $Z$ is a ${\cal D}$%
-covering of $Y$, i.e. $Y=Z/{\cal D}$. The set of maps $\left( i_a\right) $
that send elements $\left( d,y\right) \in {\cal D}\times V_a$ to the
equivalence classes $i_a\left( d,y\right) $ to which they belong provides
the natural collection of trivializations for this ${\cal D}$-space; it
follows that $i_b^{-1}\circ i_a\left( d,y\right) =\left( d\cdot
g_{ab},y\right) $, so that the associated \u Cech cocycle is the one we have
started with. Choose base points $z\in Z$, $y=q\left( z\right) $. Now
observe that the group ${\cal D}\simeq \pi _1\left( Y,y\right) $ enters this
construction in two different ways: Firstly, the elements of ${\cal D}$
locally label the different sheets of the covering in a trivialization.
Secondly, ${\cal D}$ is the set of homotopy classes of loops at $y$ on the
base manifold $Y$. We now construct a homomorphism $\rho $ from ${\cal D}$
as the set of homotopic loops to ${\cal D}$ as the labelling space for the
sheets of the covering: Choose a homotopy class $\left[ \gamma \right] $,
where $\gamma $ is a loop at the base point $y\in Y$. The unit interval $%
\left[ 0,1\right] $ can be divided \cite{Fulton} into $0=t_0<t_1<\cdots
<t_n=1$ such that the image of each interval $\left[ t_{i-1},t_i\right] $
lies in the open set $V_{a\left( i\right) }$. Then every point $\gamma
\left( t_i\right) $ lies in $V_{a\left( i\right) }\cap V_{a\left( i+1\right)
}$; on this domain, the cocycle $g_{a\left( i\right) a\left( i+1\right) }$
is constant. We lift the loop $\gamma $ to a curve $\tilde \gamma $ starting
at the base point $z$ in $Z$. If $i_{a\left( 0\right) }\left( d,y\right)
=z=\tilde \gamma \left( 0\right) $, then we find that 
\begin{equation}
\label{pp6t1muVaFu12}i_{a\left( 0\right) }^{-1}\tilde \gamma \left( 1\right)
=\left( d\cdot g_{a\left( 0\right) a\left( 1\right) }\cdot \cdots \cdot
g_{a\left( n\right) a\left( 0\right) },y\right) \equiv \left( d\cdot \rho
\left[ \gamma \right] ,y\right) \quad , 
\end{equation}
which defines an element $\rho \left[ \gamma \right] \equiv g_{a\left(
0\right) a\left( 1\right) }\cdot \cdots \cdot g_{a\left( n\right) a\left(
0\right) }\in {\cal D}$. It can be shown that this is independent of the
representative $\gamma $ of the homotopy class $\left[ \gamma \right] $, and
furthermore, that the assignment $\left[ \gamma \right] \mapsto \rho \left[
\gamma \right] $ is a homomorphism.

Thus, ${\cal D}$-coverings of $Y$, or equivalently, \u Cech cocycles on $%
{\cal V}$, are characterized by, and in turn characterize, homomorphisms $%
{\cal D}\rightarrow {\cal D}$. Cohomologous cocycles $g_{ab}^{\prime
}=h_a^{-1}g_{ab}h_b$ give rise to homomorphisms $\rho $, $\rho ^{\prime }$
that differ by conjugation, i.e. an inner automorphism of ${\cal D}$, $\rho
^{\prime }\left[ \gamma \right] =h_{a\left( 0\right) }^{-1}\cdot \rho \left[
\gamma \right] \cdot h_{a\left( 0\right) }$. For example, the homomorphism $%
{\cal D}\rightarrow {\cal D}$ is the trivial one, i.e. $\left[ \gamma
\right] \mapsto e\in {\cal D}$ for all elements $\left[ \gamma \right] $ in $%
{\cal D}$, if and only if the associated ${\cal D}$-covering of $Y$ is
(isomorphic to) the trivial $\#{\cal D}$-sheeted covering ${\cal D}\times
Y\rightarrow Y$ of $Y$, ${\cal D}$ acting on ${\cal D}\times Y$ by left
multiplication on the first factor. On the other hand, we now show that the
class of ${\cal D}$-isomorphic simply connected ${\cal D}$-coverings $%
p:X\rightarrow Y=X/{\cal D}$ is characterized by homomorphisms $\rho $ which
are inner automorphisms of ${\cal D}$: To see this, we first examine the
simply connected covering space $X$ consisting of all homotopy classes $%
\left[ \gamma \right] $ of curves $\gamma $ in $Y$ with initial point $%
\gamma \left( 0\right) =y$, where $y$ is the base point of $Y$. Choose
trivializations $i_a$ on ${\cal D}\times V_a$ so that the image $%
i_a^{-1}\left[ c\right] $ of the constant loop $c$ at $y$ is represented by $%
\left( \left[ c\right] ,y\right) $ for all $a$ for which $y\in V_a$. Then an
arbitrary loop class $\left[ \gamma \right] \in \pi _1\left( Y,y\right) $ is
represented by $\left[ \gamma \right] \equiv i_a\left( \left[ \gamma \right]
,y\right) \in X$; but this element is just the endpoint $\tilde \gamma
\left( 1\right) $ of the lift $\tilde \gamma $ of $\gamma $ to $\left[
c\right] \in p^{-1}\left( y\right) $, which implies by formula (\ref
{pp6t1muVaFu12}) that $\rho \left[ \gamma \right] =\left[ \gamma \right] $.
Since this holds for all $\left[ \gamma \right] $, we have $\rho =\left.
id\right| {\cal D}$ in this case. Now, if $p^{\prime }:X^{\prime
}\rightarrow Y$ is another simply connected covering, we have seen above
that the associated cocycles are cohomologous, hence the associated ${\cal D}
$-homomorphisms differ by an inner automorphism; but since $\rho $ is the
identity, this means that every homomorphism $\rho ^{\prime }:{\cal D}%
\rightarrow {\cal D}$ must be an inner automorphism of ${\cal D}$.

The developments of the last paragraph together with the content of theorems 
\ref{Sheets} and \ref{Cozykel} are summarized in

\subsection{Theorem: Multi-valued potentials \label{MultiValuedPotential}}

Let $\alpha $ be a closed $1$-form on the smooth manifold $Y$ with base
point $y$. Let ${\cal V}=\left\{ V_a\mid a\in A\right\} $ be a simply
connected path-connected open cover of $Y$. Let ${\cal D}\equiv \pi _1\left(
Y,y\right) $. Then

\begin{description}
\item[(A)]  \quad for every ${\cal D}$-valued $1$-\u Cech-cocycle $\left(
g_{ab}\right) $, $a,b\in A$, on ${\cal V}$ whose associated homomorphism $%
\rho :{\cal D}\rightarrow {\cal D}$ is an {\it inner} automorphism of ${\cal %
D}$, i.e. $\rho \left( d^{\prime }\right) =d\cdot d^{\prime }\cdot d^{-1}$
for some fixed $d\in {\cal D}$, there exists a collection of functions $%
f_{a,d}:V_a\rightarrow \TeXButton{R}{\mathbb{R}}$ for $a\in A$, $d\in {\cal D%
}$, such that

\begin{enumerate}
\item  \quad $f_{a,d}$ is a local potential for $\alpha $, i.e. $%
df_{a,d}=\alpha $ on $V_a$, for all $a\in A$ and $d\in {\cal D}$;

\item  \quad let $\lambda $ be a loop at $y$ with $\left[ \lambda \right]
=d\in \pi _1\left( Y,y\right) \simeq {\cal D}$. Then 
\begin{equation}
\label{pp6t1SheetsOfPotential}f_{a,d}=f_{a,e}+\int\limits_\lambda \alpha
\quad , 
\end{equation}
where $e$ is the identity in ${\cal D}$.

\item  \quad the $f_{a,d}$ satisfy a {\it glueing condition}, expressed by 
\begin{equation}
\label{pp6t1MultiValFunc1}f_{a,d}=f_{b,d\cdot g_{ab}} 
\end{equation}
on $V_a\cap V_b\neq \emptyset $.
\end{enumerate}

\item[(B)]  \quad Let $\left( g_{ab}^{\prime }\right) $ be a cocycle
cohomologous to $\left( g_{ab}\right) $, and let $\left( f_{a,d}^{\prime
}\right) $ be a collection of functions on ${\cal V}$ satisfying properties
(A1--A3) with respect to $\left( g_{ab}^{\prime }\right) $. Then there
exists a real constant $c$ and a ${\cal D}$-valued $0$-\u Cech cochain $%
\left( k_a:V_a\rightarrow {\cal D}\right) $ on ${\cal V}$ such that 
\begin{equation}
\label{pp6t1MultiValFunc2}f_{a,d}^{\prime }=f_{a,d\cdot k_a}+c 
\end{equation}
for all $a\in A$, $d\in {\cal D}$. The $0$-cochain $\left( k_a\right) $ is
determined by the cocycles $\left( g_{ab}\right) $ and $\left(
g_{ab}^{\prime }\right) $ up to its value $k_{a_0}$ on the open set $%
V_{a_0}\in {\cal V}$ which contains the base point $y$; on $V_{a_0}$, $%
k_{a_0}$ can range arbitrarily in the coset $h_{a_0}^{-1}\cdot {\cal D}%
_{center}$, where $h_{a_0}$ is the value of the \u Cech cochain which
relates the cocycles $\left( g_{ab}\right) $ and $\left( g_{ab}^{\prime
}\right) $ on $V_{a_0}$, and ${\cal D}_{center}$ is the center of ${\cal D}$.

\item[(C)]  \quad {\bf Definition:\quad }A collection $\left(
g_{ab};f_{a,d}\right) $ satisfying properties (A1--A3) will be called a {\it %
multi-valued potential function} for the closed $1$-form $\alpha $ on $Y$.
\end{description}

\TeXButton{Beweis}{\raisebox{-1ex}{\it Proof :}
\vspace{1ex}}

\underline{Ad (A) :}\quad Let $X$ be the identification space $i_a:{\cal D}%
\times V_a\rightarrow X\equiv \bigsqcup\limits_{a\in A}{\cal D}\times
V_a/\sim $, where the relation $\left( d,y\right) \sim \left( d^{\prime
},y^{\prime }\right) $ for $\left( d,y\right) \in {\cal D}\times V_a$, $%
\left( d^{\prime },y^{\prime }\right) \in {\cal D}\times V_b$ is defined to
be true if and only if $\left( d^{\prime },y^{\prime }\right) =\left( d\cdot
g_{ab},y\right) $, in which case these elements are identified according to $%
i_a\left( d,y\right) =i_b\left( d^{\prime },y^{\prime }\right) $. Then $X$
is a covering space of $Y$, and $Y$ is the space of orbits on $X$ under the
action of ${\cal D}$ on $X$ according to $\left( d^{\prime },\left(
d,y\right) \right) \mapsto \left( d^{\prime }\cdot d,y\right) $. Since the
homomorphism $\rho :{\cal D\rightarrow D}$ associated with $\left(
g_{ab}\right) $ is an inner automorphims by assumption, it follows from the
discussion at the beginning of this section that $X$ is simply connected.
Since $Y$ is a smooth manifold, $X$ is a smooth manifold. Let $%
p:X\rightarrow Y$ be the projection, which is a local diffeomorphism. Then $%
p^{*}\alpha $ is a closed, hence exact, $1$-form on $X$, and has a potential 
$F$ with $dF=p^{*}\alpha $. The identification maps $i_a:{\cal D}\times
V_a\rightarrow p^{-1}\left( V_a\right) $ are the natural trivializations for
this covering. If we write $i_a\left( d,y\right) \equiv i_{a,d}\left(
y\right) $, then $i_{a,d}$ is the inverse of the restriction $\left.
p\right| i_a\left( \left\{ d\right\} \times V_a\right) $. Now define $%
f_{a,d}\left( y\right) \equiv F\circ i_a\left( d,y\right) $ for $y\in V_a$.
By construction, $df_{a,d}=i_{a,d}^{*}dF$, and since $dF=\left[ \left.
p\right| i_a\left( \left\{ d\right\} \times V_a\right) \right] ^{*}\alpha $,
it follows that $df_{a,d}=\alpha $ on $V_a$. Furthermore, if also $y\in V_b$%
, then $F\circ i_a\left( d,y\right) =\left( F\circ i_b\right) \circ \left(
i_b^{-1}\circ i_a\right) \left( d,y\right) =\left( F\circ i_b\right) \left(
d\cdot g_{ab},y\right) =f_{b,d\cdot g_{ab}}\left( y\right) $. Formula (\ref
{pp6t1SheetsOfPotential}) is a consequence of theorems \ref{Sheets} and \ref
{Cozykel}. This proves (A).

\underline{Ad (B) :}\quad From the cocycles $\left( g_{ab}\right) $, $\left(
g_{ab}^{\prime }\right) $, construct coverings $p:X\rightarrow Y$, $%
q:Z\rightarrow Y$ as in the proof of (A), with trivializations $i_a:{\cal D}%
\times V_a\rightarrow X$, $j_a:{\cal D}\times V_a\rightarrow Z$. Then both $%
X $ and $Z$ are smooth, simply connected manifolds. The glueing condition $%
f_{a,d}=f_{b,d\cdot g_{ab}}$ for $\left( f_{a,d}\right) $ implies that there
exists a smooth function $F$ on $X$ such that $F\circ i_{a,d}=f_{a,d}$: For,
we have $f_{a,d}\left( y\right) =f_{b,d^{\prime }}\left( y^{\prime }\right) $
whenever $i_a\left( d,y\right) =i_b\left( d^{\prime },y^{\prime }\right) $;
the universal property of the identification space \cite{Brown} $i_a:{\cal D}%
\times V_a\rightarrow X$ guarantees the existence of a smooth $F$ with the
desired property. A similar function $F^{\prime }$ with $f^{\prime }\circ
j_{a,d}=f_{a,d}^{\prime }$ exists on $Z$. Since $X$ and $Z$ are ${\cal D}$%
-isomorphic, there exists a diffeomorphism $\phi :Z\rightarrow X$ preserving
fibres, i.e. $p\circ \phi =q$, and being ${\cal D}$-equivariant, i.e. $\phi
\left( d\cdot z\right) =d\cdot \phi \left( z\right) $. In the
trivializations employed above we have 
\begin{equation}
\label{pp6t1zwi20}i_a^{-1}\circ \phi \circ j_a\left( d,y\right) =\left(
d\cdot k_a,y\right) \quad , 
\end{equation}
where the collection $\left( k_a:V_a\rightarrow {\cal D}\right) $ defines a $%
0$-\u Cech cochain on ${\cal V}$.

Since $d\left( F\circ \phi \right) =\phi ^{*}dF=\phi ^{*}p^{*}\alpha
=q^{*}\alpha $, we see that both $F\circ \phi $ and $F^{\prime }$ are
potentials for $q^{*}\alpha $ on $Z$; since $Z$ is simply connected, it
follows that $F^{\prime }=F\circ \phi +c$ with $c\in \TeXButton{R}
{\mathbb{R}}$. Then%
$$
f_{a,d}^{\prime }\left( y\right) =F^{\prime }\circ j_a\left( d,y\right)
=F\circ \left( \phi \circ j_a\right) \left( d,y\right) +c=\left( F\circ
i_a\right) \left( d\cdot k_a,y\right) +c=f_{a,d\cdot k_a}\left( y\right)
+c\quad , 
$$
where we have used (\ref{pp6t1zwi20}); thus (\ref{pp6t1MultiValFunc2})
follows.

Furthermore, assume that $V_a\cap V_b\neq \emptyset $, then the analogue of (%
\ref{pp6t1zwi20}) on $V_b$ reads $i_b^{-1}\circ \phi \circ j_b\left( d\cdot
g_{ab}^{\prime },y\right) =\left( d\cdot g_{ab}^{\prime }\cdot k_b,y\right) $%
, which implies%
$$
\left( i_b^{-1}\circ i_a\right) \circ \left( i_a^{-1}\circ \phi \circ
j_a\right) \circ \left( j_a^{-1}\circ j_b\right) \left( d\cdot
g_{ab}^{\prime },y\right) =\left( d\cdot g_{ab}^{\prime }\cdot k_b,y\right)
\quad , 
$$
from which it follows that 
\begin{equation}
\label{pp6t1zwi40}g_{ab}^{\prime }\cdot k_b=k_a\cdot g_{ab}\quad . 
\end{equation}
This formula says that the cochain $\left( k_a\right) $ is not arbitrary,
but is completely determined by the cocycles $\left( g_{ab}\right) $ and $%
\left( g_{ab}^{\prime }\right) $, once a choice has been made for the value
of $k_a$ on one selected $V_a$ (e.g. the $V_{a_0}$ which contains the base
point $y$ of $Y$). This follows from path-connectedness of $Y$; for, if $%
V_{a_n}$ is any open set in ${\cal V}$, there exist finite sequences $%
g_{a_0a_1}^{\prime }$, $\ldots $, $g_{a_{n-1}a_n}^{\prime }$, and $%
g_{a_0a_1} $, $\ldots $, $g_{a_{n-1}a_n}$ so that%
$$
k_{a_n}=g_{a_{n-1}a_n}^{\prime -1}\cdots g_{a_0a_1}^{\prime -1}\cdot
k_{a_0}\cdot g_{a_0a_1}\cdots g_{a_{n-1}a_n}\quad . 
$$
Since the selected $k_{a_0}$ is constant on $V_{a_0}$, a choice of $k_{a_0}$
is just a choice of an element of ${\cal D}$. Furthermore, if $l$ is any
loop in $Y$ at the base point $y$ representing the element $\delta \in {\cal %
D}$, then from (\ref{pp6t1muVaFu12}) we see that the associated series $%
\left( g_{a_ia_{i+1}}\right) $ and $\left( g_{a_ia_{i+1}}^{\prime }\right) $
represent $\rho \left[ l\right] \equiv g_{a\left( 0\right) a\left( 1\right)
}\cdots g_{a\left( n\right) a\left( 0\right) }$, $\rho ^{\prime }\left[
l\right] \equiv g_{a\left( 0\right) a\left( 1\right) }^{\prime }\cdots
g_{a\left( n\right) a\left( 0\right) }^{\prime }$, respectively. Thus, we
must have%
$$
k_{a_0}=\rho ^{\prime }\left[ l\right] ^{-1}\cdot k_{a_0}\cdot \rho \left[
l\right] 
$$
for all homotopy classes $\left[ l\right] \in \pi _1\left( Y,y\right) \equiv 
{\cal D}$. However, as ${\cal D}$ acts transitively on each fibre of the
covering $p:X\rightarrow Y$, it follows that the elements $\rho ^{\prime
}\left[ l\right] $, $\rho \left[ l\right] $ take any value in ${\cal D}$, as 
$\left[ l\right] $ ranges in ${\cal D}$. Using $\rho ^{\prime }\left[
l\right] =h_{a_0}^{-1}\cdot \rho \left[ l\right] \cdot h_{a_0}$ it follows
that $k_{a_0}=h_{a_0}^{-1}\cdot \rho \left[ l\right] ^{-1}\cdot h_{a_0}\cdot
k_{a_0}\cdot \rho \left[ l\right] $, or%
$$
\delta \cdot \left( h_{a_0}k_{a_0}\right) =\left( h_{a_0}k_{a_0}\right)
\cdot \delta 
$$
for all $\delta \in {\cal D}$. But this implies that $h_{a_0}k_{a_0}$ must
lie in the center of ${\cal D}$, which is a normal subgroup of ${\cal D}$.
Therefore $k_{a_0}$ can range in the coset $h_{a_0}^{-1}\cdot {\cal D}%
_{center}$. Hence the collection $\left( f_{a,d}\right) $ is determined up
to a real constant and an arbitrary element in $h_{a_0}^{-1}\cdot {\cal D}%
_{center}$. \TeXButton{BWE}{\hfill
\vspace{2ex}
$\blacksquare$}

\TeXButton{Abst}{\vspace{0.8ex}}In the sequel we apply the covering
techniques discussed so far to coverings of symplectic manifolds. We first
present our notational conventions:

\section{Notation and conventions for symplectic manifolds \label
{NotationSymplecticManifolds}}

Sections \ref{NotationSymplecticManifolds}--\ref{HamiltonianVectorFields}
are based on \cite{Woodhouse}, \cite{GuillStern}, \cite{CramPir}.

--- We recall that a symplectic form $\omega $ on a manifold $M$ is a
closed, nondegenerate $2$-form on $M$. In this case, the pair $\left(
M,\omega \right) $ is called a symplectic manifold.

--- By ${\cal F}\left( M\right) $ we denote the set of all smooth functions $%
f:M\rightarrow \TeXButton{R}{\mathbb{R}}$. On a symplectic manifold we can
make ${\cal F}\left( M\right) $ into a real Lie algebra using Poisson
brackets.

--- By $\chi \left( M\right) $ we denote the set of all smooth vector fields
on $M$.

--- If $G$ is a Lie group, we will frequently denote its Lie algebra by $%
\hat g$, and the coalgebra, i.e. the space dual to $\hat g$, by $g^{*}$.

--- Given an action $\phi :G\times M\rightarrow M$ of a Lie group $G$ on a
manifold $M$, we will frequently denote the components of its tangent map $%
\phi _{*}$ by $\phi _{*}=\left( \frac{\partial \phi }{\partial G},\frac{%
\partial \phi }{\partial M}\right) $. If $A\in \hat g$, the induced vector
field on $M$ will be denoted by $\frac{\partial \phi }{\partial G}A$ or $%
\tilde A$.

--- Interior multiplication of a vector $V$ with a $q$-form $\omega $ will
be denoted by $V\TeXButton{i}{\intmul}\omega $.

--- A diffeomorphism $f:M\rightarrow M$ on a symplectic manifold $\left(
M,\omega \right) $ is called canonical transformation, if $f^{*}\omega
=\omega $. An action $\phi :G\times M\rightarrow M$ of a Lie group $G$ on $M$
is called symplectic if every $\phi _g$ is a canonical transformation.

--- $I$ generally denotes the closed interval $I=\left[ 0,1\right] \subset 
\TeXButton{R}{\mathbb{R}}$.

\section{Hamiltonian and locally Hamiltonian vector fields \label
{HamiltonianVectorFields}}

On a symplectic manifold, the symplectic form $\omega $ provides a
non-natural isomorphism between tangent spaces $T_xM$ and cotangent spaces $%
T_x^{*}M$ at every point $x\in M$, since $\omega $ is non-degenerate. In
particular, for every $f\in {\cal F}\left( M\right) $ there exists a unique
vector field $\rho f\in \chi \left( M\right) $ such that 
\begin{equation}
\label{pp6t1fo1}\rho f\TeXButton{i}{\intmul}\omega +df=0\quad . 
\end{equation}
This gives us a well-defined map $\rho :{\cal F}\left( M\right) \rightarrow
\chi \left( M\right) $. A vector field $V\in \chi \left( M\right) $ which is
the image of a function $f\in {\cal F}\left( M\right) $ under $\rho $, $%
V=\rho f$, is called Hamiltonian. The set of all (smooth) Hamiltonian vector
fields on $M$ is denoted by $\chi _H\left( M\right) $, and is a real vector
space.

On the other hand, the set $\chi _{LH}\left( M\right) $ of vector fields $V$
on $M$ which satisfy 
\begin{equation}
\label{pp6t1fo2}{\cal L}_V\omega =0\quad , 
\end{equation}
where ${\cal L}_V$ denotes a Lie derivative, is called the set of locally
Hamiltonian vector fields. This means that on every simply connected open
neighbourhood $U\subset M$, the $1$-form $V\TeXButton{i}{\intmul}\omega $ is
exact, hence there exists a smooth function $f\in {\cal F}\left( U\right) $
such that 
\begin{equation}
\label{pp6t1fo4}V\TeXButton{i}{\intmul}\omega +df=0\quad \text{on }U\quad . 
\end{equation}
As (\ref{pp6t1fo4}) holds in a neighbourhood of every point, we refer to $V$
as a locally Hamiltonian vector field. The functions $f\in {\cal F}\left(
U\right) $ need not be globally defined. If $M$ is simply connected, then
every locally Hamiltonian vector field is Hamiltonian, and $\chi _{LH}\left(
M\right) =\chi _H\left( M\right) $. This will not be true for the manifolds
we are interested in in this work.

\section{Cotangent bundles \label{CotangentBundles}}

In this section we compile some standard facts about cotangent bundles we
shall use throughout this paper. This material is discussed in standard
textbooks on Symplectic Geometry (e.g. \cite{GuillStern}), Mechanics (e.g. 
\cite{AbrahamMarsden}), Differential Geometry (e.g. \cite{CramPir}), and
Algebraic Topology (e.g. \cite{DodsonParker}).

--- On the cotangent bundle $T^{*}M$ of a manifold $M$ we have a projection $%
\tau :T^{*}M\rightarrow M$, and a natural symplectic $2$-form being given as
the differential $\omega =d\theta $ of the canonical $1$-form \cite
{GuillStern}, \cite{CramPir} $\theta $ on $T^{*}M$, which is defined as
follows: For $V\in T_{\left( m,p\right) }T^{*}M$, the action of $\theta $ on 
$V$ is defined by $\left\langle \theta ,V\right\rangle \left( m,p\right)
\equiv \left\langle p,\tau _{*}V\right\rangle $.

--- The homotopy groups of $T^{*}M$ are determined by those of $M$; in fact
we have 
\begin{equation}
\label{pp6t1homotopyGroups}\pi _n\left( T^{*}M\right) \simeq \pi _n\left(
M\right) 
\end{equation}
for all $n\ge 0$. This follows from the exact homotopy sequence for
fibrations (see any textbook on Algebraic Topology, e.g. \cite{DodsonParker}%
), 
\begin{equation}
\label{pp6t1homSeq}\cdots \rightarrow \pi _{n+1}\left( B\right) \stackrel{%
\partial }{\longrightarrow }\pi _n\left( F\right) \stackrel{i_{\#}}{%
\longrightarrow }\pi _n\left( E\right) \stackrel{p_{\#}}{\longrightarrow }%
\pi _n\left( B\right) \rightarrow \cdots \quad , 
\end{equation}
where $E\stackrel{p}{\longrightarrow }B$ is a fibration with standard fibre $%
F$. For a vector bundle with $F\simeq \TeXButton{R}{\mathbb{R}}^k$, the
homotopy groups $\pi _n\left( \TeXButton{R}{\mathbb{R}}^k\right) $ are
trivial, hence%
$$
0\rightarrow \pi _n\left( E\right) \stackrel{p_{\#}}{\longrightarrow }\pi
_n\left( B\right) \rightarrow 0 
$$
is an exact sequence, which says that $p_{\#}$ is an isomorphism in this
case. As a consequence, (\ref{pp6t1homotopyGroups}) follows.

A diffeomorphism $f:M\rightarrow M$ can be extended to a diffeomorphism $%
^{*}f:T^{*}M\rightarrow T^{*}M$ as follows \cite{GuillStern}: For $\left(
m,p\right) \in T^{*}M$, let 
\begin{equation}
\label{pp6t1cot1}\left( ^{*}f\right) \left( m,p\right) \equiv \left( f\left(
m\right) ,\left( f^{-1}\right) ^{*}p\right) \quad . 
\end{equation}
$^{*}f$ is fibre-preserving and hence a bundle map. Given two
diffeomorphisms $f,g$ we find 
\begin{equation}
\label{pp6t1cot1a}^{*}\left( fg\right) =\left( ^{*}f\right) \left(
^{*}g\right) \quad . 
\end{equation}
From definition (\ref{pp6t1cot1}) it follows immediately that every $^{*}f$
preserves the canonical $1$-form $\theta $, 
\begin{equation}
\label{pp6t1cot3}\left( ^{*}f\right) ^{*}\theta =\theta \quad . 
\end{equation}

\section{Coverings of cotangent bundles \label{CoveringCotBundle}}

In this section we start with symplectic manifolds $\left( T^{*}Y,d\theta
\right) $ which are the cotangent bundles $T^{*}Y$ of non-simply connected
configuration spaces $Y$, and are endowed with the natural symplectic $2$%
-form $d\theta $ which is associated with the canonical $1$-form $\theta $
on the cotangent bundle. Then we extend a simply connected covering manifold 
$X$ of $Y$ to a covering manifold $T^{*}X$ of $T^{*}Y$ and show how the
projection map $p:X\rightarrow Y$ can be extended to give a local
symplectomorphism between $T^{*}X$ and $T^{*}Y$ which is also a covering map.

\subsection{Covering spaces and their cotangent bundles}

Let $p:X\rightarrow Y$ be a covering of manifolds. Let $\Theta ,\theta $
denote the canonical $1$-forms on the cotangent bundles $T^{*}X$, $T^{*}Y$,
respectively (see section \ref{CotangentBundles}). The bundle projections
are written as $\sigma :T^{*}X\rightarrow X$ and $\tau :T^{*}Y\rightarrow Y$%
; the canonical symplectic $2$-forms on $T^{*}X$, $T^{*}Y$ are $\Omega
=d\Theta $, $\omega =d\theta $, respectively.

Central to our developments is the observation that we can extend the
projection $p$ to a covering map of cotangent bundles; for all coverings it
is understood that they are smooth:

\subsection{Theorem \label{PStarIsACovering}}

Let $p:X\rightarrow Y$ be a covering. Then $p$ can be extended to a bundle
map $^{*}p:T^{*}X\rightarrow T^{*}Y$ such that $^{*}p:T^{*}X\rightarrow
T^{*}Y$ is a covering.

\TeXButton{Beweis}{\raisebox{-1ex}{\it Proof :}
\vspace{1ex}}

We first have to show that a well-defined extension $^{*}p$ exists. To this
end, let $V\subset Y$ be an admissible open neighbourhood in $Y$, and let $%
U\subset X$ denote a connected component of $p^{-1}\left( V\right) $. On
every $U$, the restriction $\left. p\right| U:U\rightarrow Y$ is a
diffeomorphism, hence its extension 
\begin{equation}
\label{pp6t1pstarisacovering}^{*}\left( \left. p\right| U\right) \equiv
\left( \left. p\right| U,\left[ \left. p\right| U\right] ^{-1*}\right) 
\end{equation}
is well-defined. The collection of all $U$, being the collection of all
inverse images of admissible open neighbourhoods $V$ in $Y$ forms an open
cover of $X$. On the intersection of any two of the $U,U^{\prime }$, the
locally defined maps (\ref{pp6t1pstarisacovering}) coincide. Now define $%
^{*}p$ to be the uniquely determined function on $T^{*}X$ whose restriction
to any of the $U$ coincides with $^{*}\left( \left. p\right| U\right) $.

From this it follows that $^{*}p$ is well-defined, and is smooth, provided $%
p $ is smooth. Its local form (\ref{pp6t1pstarisacovering}) shows that $%
^{*}p $ preserves fibres and hence is a bundle map. If $V$ is any admissible
neighbourhood in $Y$, then $\bigcup\limits_{y\in V}T_y^{*}Y$ is a
neighbourhood in $T^{*}Y$ whose inverse image under $^{*}p$ is a disjoint
union of neighbourhoods in $T^{*}X$. This says that $^{*}p$ is a covering
map. \TeXButton{BWE}{\hfill
\vspace{2ex}
$\blacksquare$}

\TeXButton{Abst}{\vspace{0.8ex}}As a consequence of the last theorem, we can
pull back $q$-forms $\omega $ on $T^{*}Y$ to $q$-forms $\left( ^{*}p\right)
^{*}\omega $ on $T^{*}X$ via $^{*}p$. Hence the pull-back $\left(
^{*}p\right) ^{*}\theta $ is well-defined. The next theorem explains the
relation between this pull-back and the canonical symplectic potential $%
\Theta $ on $T^{*}X$. To this end we note that

\subsection{Lemma}

\begin{equation}
\label{pp6t1cover2}\tau \circ \left( ^{*}p\right) =p\circ \sigma \quad . 
\end{equation}
\TeXButton{Beweis}{\raisebox{-1ex}{\it Proof :}
\vspace{1ex}}

This follows from the definition $\left( ^{*}p\right) \left( x,\alpha
\right) =\left( p\left( x\right) ,\left( \left. p\right| U\right)
^{-1*}\alpha \right) $ for $\left( x,\alpha \right) \in T^{*}X$. 
\TeXButton{BWE}{\hfill
\vspace{2ex}
$\blacksquare$}

\subsection{Theorem}

The pull-back of the canonical $1$-form $\theta $ on $Y$ under $^{*}p$
coincides with the canonical $1$-form $\Theta $ on $X$,

\begin{equation}
\label{pp6t1cover3}\left( ^{*}p\right) ^{*}\theta =\Theta \quad . 
\end{equation}
\TeXButton{Beweis}{\raisebox{-1ex}{\it Proof :}
\vspace{1ex}}

Let $\left( x,\alpha \right) \in T^{*}X$, let $V\in T_{\left( x,\alpha
\right) }T^{*}X$. Then $\left( ^{*}p\right) \left( x,\alpha \right) =\left(
p\left( x\right) ,\left( \left. p\right| U\right) ^{-1*}\alpha \right) $,
where $U$ is a neighbourhood of $x$ in $X$ on which the restriction of $p$
is injective. Now if $\left( ^{*}p\right) _{*}V$ is paired with $\theta $ at
the point $\left( ^{*}p\right) \left( x,\alpha \right) \in T^{*}Y$, we
obtain $\left\langle \theta ,\left( ^{*}p\right) _{*}V\right\rangle
=\left\langle \left( \left. p\right| U\right) ^{-1*}\alpha ,\tau _{*}\left(
^{*}p\right) _{*}V\right\rangle $; however, from (\ref{pp6t1cover2}) we have
that $\tau _{*}\left( ^{*}p\right) _{*}=p_{*}\sigma _{*}$, and hence%
$$
\left\langle \left( \left. p\right| U\right) ^{-1*}\alpha ,\tau _{*}\left(
^{*}p\right) _{*}V\right\rangle =\left\langle \alpha ,\sigma
_{*}V\right\rangle =\left\langle \Theta ,V\right\rangle \quad , 
$$
by the definition of $\Theta $; this proves the theorem. \TeXButton{BWE}
{\hfill
\vspace{2ex}
$\blacksquare$}

\TeXButton{Abst}{\vspace{0.8ex}}There is an immediate important consequence:

\subsection{Corollary \label{LocalSymplectomorphism}}

The canonical symplectic $2$-form $d\Theta $ on $T^{*}X$ is the pull-back of
the canonical symplectic $2$-form $d\theta $ on $T^{*}Y$ under $^{*}p$, and
hence the map $^{*}p:T^{*}X\rightarrow T^{*}Y$ is a {\bf local
symplectomorphism}.

\TeXButton{Beweis}{\raisebox{-1ex}{\it Proof :}
\vspace{1ex}}

This follows from $d\Theta =d\left( ^{*}p\right) \theta =\left( ^{*}p\right)
d\theta $. \TeXButton{BWE}{\hfill
\vspace{2ex}
$\blacksquare$}

\TeXButton{Abst}{\vspace{0.8ex}}Hence we can relate, and locally identify,
the dynamics taking place on $T^{*}Y$ to an associated dynamical system on
the symplectic covering manifold $T^{*}X$; this is one of the major
statements of this work. As was shown in (\ref{pp6t1homotopyGroups}), the
homotopy groups of a cotangent bundle are isomorphic to those of its base
space. In particular, if $X$ is simply connected, the fundamental groups
obey the relations 
\begin{equation}
\label{pp6t1fndGrp}\pi _1\left( T^{*}X\right) =\pi _1\left( X\right) =0\quad
,\quad \pi _1\left( T^{*}Y\right) =\pi _1\left( Y\right) ={\cal D}\quad , 
\end{equation}
and it follows that the deck transformation group ${\cal D}\left(
T^{*}X\right) $ of the covering $^{*}p:T^{*}X\rightarrow T^{*}Y$ is just $%
\pi _1\left( T^{*}Y\right) ={\cal D}$. This will enable us to remove the
multi-valuedness of a local moment map given on a cotangent space $T^{*}Y$
by constructing the local symplectomorphism $^{*}p$, and then studying the
associated dynamics on the simply connected symplectic covering space $%
T^{*}X $, on which every locally Hamiltonian vector field has a globally
defined charge, and hence every symplectic group action has a global moment
map; see section \ref{LocalMomentMaps} for the details. Our next task
therefore is to study how (Lie) group actions defined on $Y$ (and, in turn, $%
T^{*}Y$) can be lifted to a covering space space $X$ (and $T^{*}X$), in
particular, when $X$ is simply connected; this is done in sections \ref
{LiftOfGroupActions} and \ref{PreservationOfTheGroupLaw} for general
symplectic manifolds which are not necessarily cotangent bundles.

\section{Lift of group actions under covering maps \label{LiftOfGroupActions}
}

In this section we prove a couple of theorems about the lifting of a left
action $\phi $ of a Lie group $G$ on a manifold $Y$ to a covering manifold $%
X $. Here we examine existence and uniqueness of lifts; in the next section
we examine under which conditions the group law of $G$ is preserved under
the lift.

Let $p:X\rightarrow Y$ be a covering, where $X$, $Y$ are connected, let $V$
be a manifold, let $f:V\rightarrow Y$. As explained in section \ref
{BasicsAboutCoverings}, one calls a map $\tilde f:V\rightarrow X$ a {\it %
lift of }$f${\it \ through }$p$, if $p\circ \tilde f=f$. For the sake of
convenience, we establish a similar phrase for a related construction which
will frequently appear in the following: If $f:Y\rightarrow Y$, we call $%
\hat f:X\rightarrow X$ a {\it lift of }$f${\it \ to }$X$ if 
\begin{equation}
\label{pp6t1lift01}p\circ \hat f=f\circ p\quad . 
\end{equation}
We enhance this condition for the case that $G$ is a group and $\phi
:G\times Y\rightarrow Y$ is a left action of $G$ on $Y$. In this case, we
call a smooth map $\hat \phi :G\times X\rightarrow X$ a {\it lift of }$\phi $%
{\it \ to }$X$ if

\begin{description}
\item[(L1)]  \quad $p\circ \hat \phi =\phi \circ \left( id_G\times p\right) $%
, and

\item[(L2)]  $\quad \hat \phi _e=id_X$,
\end{description}

where $\hat \phi _e$ denotes the map $x\mapsto \hat \phi _e\left( x\right)
\equiv \hat \phi \left( e,x\right) $.

We remark that (L1) does {\bf not} imply that $p$ is $G$-equivariant (or a $%
G $-morphism). This is because $G$-equivariance requires $X$ to be a $G$%
-space, i.e. a manifold with a smooth left {\bf action of }$G$ on it. This
in turn means that the lift $\hat \phi $ must preserve the group law of $G$,
i.e. $\widehat{\phi _{gh}}=\hat \phi _g\hat \phi _h$. We will see shortly
that this is guaranteed only if $G$ is connected. If $G$ has several
connected components, the lift can give rise to an extension $\tilde G$ of
the original group by the deck transformation group ${\cal D}$; this is
described in theorem \ref{ExtensionOfG}. The matter of equivariance of the
covering map $p$ is taken up in section \ref{GSpacesCoveringMaps}.

We note that if a map $\hat \phi $ satisfying (L1) exists, it is determined
only up to a deck transformation $\gamma $;\ for if we define $\hat \phi
^{\prime }\equiv \hat \phi \left( id_G\times \gamma \right) $, $\hat \phi
^{\prime }$ also satisfies (L1); this is why we have to impose (L2)
additionally. However, we show that if $\hat \phi $ exists, then it can
always be assumed that it satisfies (L2):

\subsection{Proposition \label{Smoothness}}

\begin{enumerate}
\item  \quad Let $\phi :G\times Y\rightarrow Y$ be a smooth left action of $%
G $ on $Y$. If a smooth map $\hat \phi :G\times X\rightarrow X$ satisfying
(L1) exists, it can always be redefined so that 
\begin{equation}
\label{pp6t1lift2}\hat \phi _e=id_X\quad . 
\end{equation}

\item  \quad Every $\hat \phi _g$ is a diffeomorphism, and the assignment $%
\left( g,x\right) \mapsto \hat \phi _g^{-1}\left( x\right) $ is smooth.
\end{enumerate}

\TeXButton{Beweis}{\raisebox{-1ex}{\it Proof :}
\vspace{1ex}}

\underline{Ad (1) :}\quad By assumption, $p\hat \phi _e=\phi _ep=p$, hence $%
\hat \phi _e$ is a deck transformation $\gamma $ of the covering. Now
redefine $\hat \phi \mapsto \hat \phi ^{\prime }=\hat \phi \left( id_G\times
\gamma ^{-1}\right) $, then $\hat \phi ^{\prime }$ satisfies $p\hat \phi
^{\prime }=\phi \left( id_G\times p\right) $ and $\hat \phi _e^{\prime
}=id_X $.

\underline{Ad (2) :}\quad The map $\hat \phi _g\hat \phi _{g-1}:X\rightarrow
X$ projects into $p$, hence is a deck transformation $\gamma $. Thus $\hat
\phi _g^{-1}=\widehat{\phi _{g^{-1}}}\circ \gamma ^{-1}$ is smooth, since $%
\gamma ^{-1}$ is smooth, hence $\hat \phi _g$ is a diffeomorphism. The
assignment $\left( g,x\right) \mapsto \hat \phi _g^{-1}\left( x\right) =%
\widehat{\phi _{g^{-1}}}\circ \gamma ^{-1}\left( x\right) $ is smooth, since
the map $G\rightarrow G$, $g\mapsto g^{-1}$ is smooth. \TeXButton{BWE}
{\hfill
\vspace{2ex}
$\blacksquare$}

\TeXButton{Abst}{\vspace{0.8ex}}In the following we assume that $X$ and $Y$
are connected manifolds, and $Y$ is not simply connected. We note that

\subsection{Remark}

The fundamental group of the Lie group $G$ coincides with the fundamental
group of its identity component, 
\begin{equation}
\label{pp6t1remark1}\pi _n\left( G\right) =\pi _n\left( G_0\right) \text{%
\quad }. 
\end{equation}
\TeXButton{Beweis}{\raisebox{-1ex}{\it Proof :}
\vspace{1ex}}

This follows from the exact homotopy sequence (\ref{pp6t1homSeq})%
$$
\cdots \rightarrow \pi _{n+1}\left( B\right) \stackrel{\partial }{%
\longrightarrow }\pi _n\left( F\right) \stackrel{i_{\#}}{\longrightarrow }%
\pi _n\left( E\right) \stackrel{p_{\#}}{\longrightarrow }\pi _n\left(
B\right) \rightarrow \cdots \quad , 
$$
for the fibration $pr:G\rightarrow G/G_0=Ds$. Put $B=Ds$, $F=G_0$, $E=G$, $%
\pi _n\left( Ds\right) =0$ to obtain (\ref{pp6t1remark1}). \TeXButton{BWE}
{\hfill
\vspace{2ex}
$\blacksquare$}

\TeXButton{Abst}{\vspace{0.8ex}}Now we derive a necessary and sufficient
condition for the existence of a lift satisfying (L1, L2) in terms of the
fundamental group $\pi _1\left( G\right) $ of $G$. To prove this, we first
need a

\subsection{Lemma \label{HomotopyLemma}}

Let $X,Y$ be topological spaces, let $\lambda \times \mu $ be a loop in $%
X\times Y$ at $\left( x,y\right) $, where $\lambda $ is a loop in $X$ at $x$%
, and $\mu $ is a loop in $Y$ at $y$. Then $\lambda \times \mu $ is
homotopic to a product of loops 
\begin{equation}
\label{pp6t1homolemma1}\lambda \times \mu \sim \left( \left\{ x\right\}
\times \mu \right) *\left( \lambda \times \left\{ y\right\} \right) \quad , 
\end{equation}
where $"\sim "$ means ''homotopic'', and $"*"$ denotes a product of paths.

\TeXButton{Beweis}{\raisebox{-1ex}{\it Proof :}
\vspace{1ex}}

We explicitly give a homotopy effecting (\ref{pp6t1homolemma1}). Define $%
h:I\times I\rightarrow X\times Y$, $\left( s,t\right) \mapsto h\left(
s,t\right) =h_s\left( t\right) $. Here $s$ labels the loops $t\mapsto
h_s\left( t\right) $, and $t$ is the loop parameter. Define 
\begin{equation}
\label{pp6t1homolemma2}h\left( s,t\right) \equiv \left\{ 
\begin{array}{ccl}
\left( x,\mu \left( \frac t{1-\frac s2}\right) \right) & \text{for} & t\in
\left[ 0,\frac s2\right) \quad . \\ 
\left( \lambda \left( \frac{t-\frac s2}{1-\frac s2}\right) ,\mu \left( \frac
t{1-\frac s2}\right) \right) & \text{for} & t\in \left[ \frac s2,1-\frac
s2\right) \quad . \\ 
\left( \lambda \left( \frac{t-\frac s2}{1-\frac s2}\right) ,y\right) & \text{%
for} & t\in \left[ 1-\frac s2,1\right] \quad . 
\end{array}
\right. 
\end{equation}
This is the required homotopy. \TeXButton{BWE}
{\hfill
\vspace{2ex}
$\blacksquare$}

\TeXButton{Abst}{\vspace{0.8ex}}Now we can turn our theorem. To this end,
consider a left action $\phi :G\times Y\rightarrow Y$, and choose $y\in Y$
fixed. By $\phi _y$ we denote the map $\phi _y:G\rightarrow Y$, $g\mapsto
\phi _y\left( g\right) \equiv \phi \left( g,y\right) $. $\phi _y$ induces
the homomorphism $\phi _{y\#}:\pi _1\left( G,e\right) \rightarrow \pi
_1\left( Y,y\right) $. Furthermore, if $x\in p^{-1}\left( y\right) $ is any
point in the fibre over $y$, then $p$ induces a map $p_{\#}:\pi _1\left(
X,x\right) \rightarrow \pi _1\left( Y,y\right) $. We now can state:

\subsection{Theorem \label{ConditionForGroupLift}}

Let $x\in X$ arbitrary, let $y=p\left( x\right) $. Let $G$ be connected.
Then the action $\phi :G\times Y\rightarrow Y$ possesses a unique lift $\hat
\phi :G\times X\rightarrow X$ satisfying (L1, L2) if and only if 
\begin{equation}
\label{pp6t1lift3}\phi _{y\#}\,\pi _1\left( G,e\right) \subset p_{\#}\,\pi
_1\left( X,x\right) \quad . 
\end{equation}
\TeXButton{Beweis}{\raisebox{-1ex}{\it Proof :}
\vspace{1ex}}

Assume a lift $\hat \phi $ exists. Then $\hat \phi $ is the unique lift of
the map $G\times X\rightarrow Y$, $\left( g,x\right) \mapsto \phi \left(
g,p\left( x\right) \right) $ through $p$ with the property $\hat \phi \left(
e,x\right) =x$. According to the ''Lifting Map Theorem'' \ref
{LiftingMapTheorem} it follows that 
\begin{equation}
\label{pp6t1lift4}\left[ \phi \left( id_G\times p\right) \right] _{\#}\,\pi
_1\left( G\times X,\left( e,x\right) \right) \subset p_{\#}\pi _1\left(
X,x\right) \quad . 
\end{equation}
Conversely, if (\ref{pp6t1lift4}) holds, then $\phi \left( id_G\times
p\right) $ lifts uniquely to $\hat \phi $ satisfying (L1, L2). Hence we need
only show the equivalence (\ref{pp6t1lift3}) $\Leftrightarrow $ (\ref
{pp6t1lift4}).

\underline{(\ref{pp6t1lift3})\ $\Leftarrow $\ (\ref{pp6t1lift4}) :}$\quad $%
Let $\lambda $ be a loop in $G$ at $e$. Then $\lambda \times \left\{
x\right\} $ is a loop in $G\times X$ at $\left( e,x\right) $, hence $%
t\mapsto \phi \left( \lambda \left( t\right) ,p\left( x\right) \right) =\phi
_y\circ \lambda \left( t\right) $ is a loop in $Y$ at $y$ giving rise to the
class%
$$
\phi _{y\#}\left[ \lambda \right] =\left[ \phi \left( id_G\times p\right)
\right] _{\#}\left[ \lambda \times \left\{ x\right\} \right] \in 
$$
$$
\in \left[ \phi \left( id_G\times p\right) \right] _{\#}\pi _1\left( G\times
X,\left( e,x\right) \right) \subset p_{\#}\pi _1\left( X,x\right) \quad , 
$$
where the last inclusion follows from assumption, and hence (\ref{pp6t1lift3}%
) follows.

\underline{(\ref{pp6t1lift3})\ $\Rightarrow $\ (\ref{pp6t1lift4}) :}$\quad $%
Let $\lambda \times \mu $ be a loop in $G\times X$ at $\left( e,x\right) $.
Then lemma \ref{HomotopyLemma} says that%
$$
\lambda \times \mu \sim \left( \left\{ e\right\} \times \mu \right) *\left(
\lambda \times \left\{ x\right\} \right) \quad , 
$$
hence%
$$
\left[ \phi \left( id_G\times p\right) \right] \lambda \times \mu \sim
\left[ \phi \left( id_G\times p\right) \right] \left( \left\{ e\right\}
\times \mu \right) *\left[ \phi \left( id_G\times p\right) \right] \left(
\lambda \times \left\{ x\right\} \right) = 
$$
$$
=\phi \left( e,p\mu \right) *\phi \left( \lambda ,p\left( x\right) \right)
=\left( p\mu \right) *\left( \phi _y\lambda \right) \quad . 
$$
But by assumption (\ref{pp6t1lift3}), there exists a loop $\rho $ in $X$ at $%
x$ such that $\phi _y\lambda \sim p\rho $. Then the last expression in the
last line above becomes%
$$
\left( p\mu \right) *\left( \phi _y\lambda \right) \sim p\left( \mu *\rho
\right) \quad , 
$$
where $\mu *\rho $ is a loop in $X$ at $x$. But this proves%
$$
\left[ \phi \left( id_G\times p\right) \right] \left[ \lambda \times \mu
\right] \in p_{\#}\,\pi _1\left( X,x\right) \quad , 
$$
where $\left[ \lambda \times \mu \right] $ is the homotopy class of $\lambda
\times \mu $, and hence (\ref{pp6t1lift4}) follows. \TeXButton{BWE}
{\hfill
\vspace{2ex}
$\blacksquare$}

\section{Preservation of the group law \label{PreservationOfTheGroupLaw}}

In this section we show that a lift $\hat \phi $ to a simply connected
covering space is not unique, if $G$ is not connected. Every lift gives rise
to an extension of the original group $G$ by the group ${\cal D}$ of deck
transformations of the covering. If ${\cal D}$ is Abelian, all these
extensions are equivalent.

\TeXButton{Abst}{\vspace{0.8ex}}We start with theorem \ref
{ConditionForGroupLift}: This gives a condition for the existence of a
smooth map $\hat \phi $ satisfying (L1,L2) under the assumption that $G$ is
connected. We now temporarily relax the last condition and allow $G$ to be a
Lie group with several connected components. In this case we simply assume
that a lift $\hat \phi $ exists. As mentioned above, the lift $\hat \phi $
need not preserve the group law on $G$, in which case it is not an action of 
$G$ on $X$. In particular, this means, that the set 
\begin{equation}
\label{pp6t1liftset}\left\{ \hat \phi _g\mid g\in G\right\} 
\end{equation}
of diffeomorphisms $\hat \phi _g:X\rightarrow X$ is no longer a group.
Clearly, we want to know under which circumstances the lift {\bf is} an
action of some group $\tilde G$. We examine this under the assumption that $%
G $ is a semidirect product $G=G_0\odot Ds$ of its identity component and a
discrete factor $Ds$, where $G/G_0\equiv Ds$, as $G_0$ is normal in $G$. In
this case, every element of $G$ has a unique expression $\left( g,\kappa
\right) $, where $g\in G_0$, $\kappa \in Ds$. The group law is 
\begin{equation}
\label{pp6t1lift5}\left( g,\kappa \right) \left( g^{\prime },\kappa ^{\prime
}\right) =\left( g\cdot a\left( \kappa \right) g^{\prime },\kappa \kappa
^{\prime }\right) \quad , 
\end{equation}
where $a\left( \kappa \right) :G_0\rightarrow G_0$ is an outer automorphism
of $G_0$, and $a:Ds\rightarrow Aut\left( G_0\right) $ is a representation of 
$Ds$ in the automorphism group of $G_0$, as $a\left( \kappa \kappa ^{\prime
}\right) =a\left( \kappa \right) \circ a\left( \kappa ^{\prime }\right) $,
and $a\left( e\right) =id$. We first need a

\subsection{Lemma \label{DeckObjects}}

Let $\gamma \in {\cal D}$, $g,g^{\prime }\in G$. Then the maps 
\begin{equation}
\label{pp6t1lift6}\hat \phi _g\circ \gamma \circ \hat \phi _g^{-1}\quad
,\quad \hat \phi _g\circ \hat \phi _{g^{\prime }}\circ \widehat{\phi
_{gg^{\prime }}}^{-1} 
\end{equation}
are deck transformations.

\TeXButton{Beweis}{\raisebox{-1ex}{\it Proof :}
\vspace{1ex}}

This follows immediately by applying $p$ and using (L1,L2). \TeXButton{BWE}
{\hfill
\vspace{2ex}
$\blacksquare$}

\TeXButton{Abst}{\vspace{0.8ex}}Thus, for a given $g\in G$, (\ref{pp6t1lift6}%
) defines a map 
\begin{equation}
\label{pp6t1lift7}b\left( g\right) :{\cal D}\rightarrow {\cal D}\quad ,\quad
\gamma \mapsto b\left( g\right) \gamma \equiv \hat \phi _g\circ \gamma \circ
\hat \phi _g^{-1}\quad , 
\end{equation}
which implies that $b\left( g\right) \in Aut\left( {\cal D}\right) $ is a $%
{\cal D}$-automorphism. $b:G\rightarrow Aut\left( {\cal D}\right) $ need not
be a representation, however! The second expression in (\ref{pp6t1lift6})
defines a map 
\begin{equation}
\label{pp6t1lift8}\Gamma :G\times G\rightarrow {\cal D}\quad ,\quad \left(
g,g^{\prime }\right) \mapsto \Gamma \left( g,g^{\prime }\right) \equiv \hat
\phi _g\circ \hat \phi _{g^{\prime }}\circ \widehat{\phi _{gg^{\prime }}}%
^{-1}\quad . 
\end{equation}
These maps determine how the group law of $G$ is changed under the lift $%
\hat \phi $; in particular (\ref{pp6t1lift8}) shows that $\Gamma $ expresses
the deviation of the lifted diffeomorphisms $\hat \phi _g$ from forming a
group isomorphic to $G$. This is the content of the next

\subsection{Theorem \label{ExtensionOfG}}

Let $\phi :G\times Y\rightarrow Y$ be a smooth action of $G$ on $Y$. Assume
that a smooth lift $\hat \phi $ satisfying (L1,L2) exists. Then

\begin{enumerate}
\item  \quad the set $\left\{ \hat \phi _g\mid g\in G\right\} $ is no longer
a group in general; instead, the lift $\hat \phi $ provides an {\it %
extension }$\tilde G${\it \ of }$G${\it \ by the deck transformation group }$%
{\cal D}$, i.e. ${\cal D}$ is normal in $\tilde G$, and $\tilde G/{\cal D}=G$%
. The extended group is given by the set 
\begin{equation}
\label{pp6t1lift9}\tilde G=\left\{ \gamma \circ \hat \phi _g\mid \gamma \in 
{\cal D\;};\;g\in G\right\} \quad . 
\end{equation}
If the elements of this set are denoted as pairs, $\gamma \circ \hat \phi
_g\leftrightarrow \left( \gamma ,g\right) $, then the group law of $\tilde G$
is given by 
\begin{equation}
\label{pp6t1lift10}\left( \gamma ,g\right) \cdot \left( \gamma ,g^{\prime
}\right) =\left( \gamma \cdot b\left( g\right) \gamma ^{\prime }\cdot \Gamma
\left( g,g^{\prime }\right) ,gg^{\prime }\right) \quad , 
\end{equation}
and inverses are 
\begin{equation}
\label{pp6t1lift10a}\left( \gamma ,g\right) ^{-1}=\left( b\left( g\right)
^{-1}\left[ \gamma ^{-1}\cdot \Gamma ^{-1}\left( g,g^{-1}\right) \right]
,g^{-1}\right) = 
\end{equation}
\begin{equation}
\label{pp6t1lift10b}=\left( \left[ b\left( g^{-1}\right) \gamma \cdot \Gamma
\left( g^{-1},g\right) \right] ^{-1},g^{-1}\right) \quad . 
\end{equation}
The group law (\ref{pp6t1lift10}) expresses the non-closure of the set $%
\left\{ \hat \phi _g\mid g\in G\right\} $ as discussed in (\ref{pp6t1liftset}%
), since now 
\begin{equation}
\label{pp6t1lift10c}\left( e,g\right) \cdot \left( e,g^{\prime }\right)
=\left( \Gamma \left( g,g^{\prime }\right) ,gg^{\prime }\right) \quad . 
\end{equation}

\item  \quad If $H,K\subset G$ are any connected subsets of $G$, then the
restrictions $\left. b\right| H$ and $\left. \Gamma \right| H\times K$ are
constant. Hence both $b$ and $\Gamma $ descend to the quotient $G/G_0=Ds$, 
\begin{equation}
\label{pp6t1lift11}b:Ds\rightarrow {\cal D}\quad ,\quad \Gamma :Ds\times
Ds\rightarrow {\cal D}\quad . 
\end{equation}

\item  \quad In particular, on the identity component $G_0$ we have $\left.
b\right| G_0=id$ and $\left. \Gamma \right| G_0\times G_0=e$. Hence, for
elements $\left( \gamma ,g\right) \in {\cal D}\times G_0$ we have the group
law 
\begin{equation}
\label{pp6t1lift12}\left( \gamma ,g\right) \cdot \left( \gamma ^{\prime
},g^{\prime }\right) =\left( \gamma \gamma ^{\prime },gg^{\prime }\right)
\quad . 
\end{equation}
As a consequence, the identity component $G_0$ can be regarded as a subgroup
of the extension $\tilde G$, and can be identified with the set of all
elements of the form $\left( e,g\right) $, $g\in G_0$, so that (cf. \ref
{pp6t1lift10c}) 
\begin{equation}
\label{pp6t1lift12a}\left( e,g\right) \cdot \left( e,g^{\prime }\right)
=\left( e,gg^{\prime }\right) \quad . 
\end{equation}
Also, (\ref{pp6t1lift12}) says that ${\cal D}$ and $G_0$ commute. As a
consequence of all that, the identity component $\tilde G_0$ of $\tilde G$
coincides (up to isomorphism) with the identity component of $G$, $\tilde
G_0=G_0$, and hence we have an isomorphism of Lie algebras 
\begin{equation}
\label{pp6t1lift13}Lie\left( \tilde G\right) =Lie\left( G\right) =\hat
g\quad . 
\end{equation}
\end{enumerate}

\TeXButton{Beweis}{\raisebox{-1ex}{\it Proof :}
\vspace{1ex}}

\underline{Ad (1) :}\quad We first must verify that the set $\tilde G$ as
defined in (\ref{pp6t1lift9}) is indeed a group of diffeomorphisms on $X$.
Since $\hat \phi _e=id_X$ by (L2), we have that $\tilde G$ contains $e\hat
\phi _e=id_X$. Next, if $\gamma ,\gamma ^{\prime }\in {\cal D}$ and $%
g,g^{\prime }\in G$, then $\gamma \hat \phi _g\gamma ^{\prime }\hat \phi
_{g^{\prime }}=\gamma \circ b\left( g\right) \gamma ^{\prime }\circ \hat
\phi _g\circ \hat \phi _{g^{\prime }}$ according to (\ref{pp6t1lift7}); but $%
\hat \phi _g\circ \hat \phi _{g^{\prime }}=\Gamma \left( g,g^{\prime
}\right) \circ \widehat{\phi _{gg^{\prime }}}$ by (\ref{pp6t1lift8}), which
gives 
\begin{equation}
\label{pp6t1lift14}\gamma \circ \hat \phi _g\circ \gamma ^{\prime }\circ
\hat \phi _{g^{\prime }}=\left[ \gamma \circ b\left( g\right) \gamma
^{\prime }\circ \Gamma \left( g,g^{\prime }\right) \right] \circ \left[ 
\widehat{\phi _{gg^{\prime }}}\right] \quad , 
\end{equation}
where the first factor in square brackets on the RHS is an element of ${\cal %
D}$, and the second factor is a lifted diffeomorphism, and therefore the LHS
is an element of the set $\tilde G$. Furthermore, when using the pair
notation $\left( \gamma ,g\right) $ for elements of $\tilde G$, formula (\ref
{pp6t1lift14}) yields the group law (\ref{pp6t1lift10}). Finally, we must
show that inverses exist in $\tilde G$; it is easy to use (\ref{pp6t1lift10}%
) to arrive at (\ref{pp6t1lift10a}) for inverses, which means that inverses
have the form $\gamma \circ \hat \phi _g$ as required. Furthermore, from (%
\ref{pp6t1lift10}) it follows that ${\cal D}$ is normal in $\tilde G$, and
that the cosets ${\cal D}\cdot \left( \gamma ,g\right) $ obey the group law
of $G$, since%
$$
\left[ {\cal D}\cdot \left( e,g\right) \right] \left[ {\cal D}\cdot \left(
e,g^{\prime }\right) \right] ={\cal D}\cdot \left[ \left( e,g\right) \left(
e,g^{\prime }\right) \right] ={\cal D}\cdot \left( \Gamma \left( g,g^{\prime
}\right) ,gg^{\prime }\right) ={\cal D}\cdot \left( e,gg^{\prime }\right)
\quad , 
$$
thus $\tilde G/{\cal D}=G$.

\underline{Ad (2+3) :}\quad Let $\gamma \in {\cal D}$ be arbitrary, and
consider the map $G\rightarrow p^{-1}\left( p\left( x\right) \right) $, $%
g\mapsto \left[ b\left( g\right) \gamma \right] x$, where the fibre $%
p^{-1}\left( p\left( x\right) \right) $ is a discrete space for arbitrary
fixed $x\in X$. From the definition (\ref{pp6t1lift7}) and proposition \ref
{Smoothness} we see that this map is smooth; since the target space is
discrete, it must therefore be constant on every connected subset of the
domain. Since $X$ is connected, every deck transformation of the covering $%
p:X\rightarrow Y$ is uniquely determined by its value at a single point $%
x\in X$. Hence $b\left( g\right) \gamma =const.$ for all $g$ within a
connected component of $G$. On the identity component $G_0$, $b\left(
g\right) =b\left( e\right) =id_{{\cal D}}$. -- A similar argument applies to 
$\Gamma $, since (\ref{pp6t1lift8}) shows that all maps involved in the
definition of $\Gamma $ are smooth in all arguments. In particular, $\Gamma
\left( e,g\right) =\Gamma \left( g,e\right) =e_{{\cal D}}$. (\ref
{pp6t1lift12}) is a consequence of (\ref{pp6t1lift10}). \TeXButton{BWE}
{\hfill
\vspace{2ex}
$\blacksquare$}

\TeXButton{Abst}{\vspace{0.8ex}}If ${\cal D}$ is Abelian, the map $\Gamma $
defined in (\ref{pp6t1lift8}) has a special significance:

\subsection{Theorem}

If ${\cal D}$ is Abelian, then $\Gamma :G\times G\rightarrow {\cal D}$
defines a $2$-cocycle in the ${\cal D}$-valued cohomology on $G$ as defined
in appendix, section \ref{DeckTransformationValuedCohomOnG}.

\TeXButton{Beweis}{\raisebox{-1ex}{\it Proof :}
\vspace{1ex}}

The proof is a standard argument: Compute $\left( \hat \phi _g\hat \phi
_h\right) \hat \phi _k=\hat \phi _g\left( \hat \phi _h\hat \phi _k\right) $
using (\ref{pp6t1lift8}); this leads to%
$$
b\left( g\right) \Gamma \left( h,k\right) +\Gamma \left( g,hk\right) -\Gamma
\left( gh,k\right) -\Gamma \left( g,h\right) =\left( \delta \Gamma \right)
\left( g,h,k\right) =0\quad , 
$$
where we have used (\ref{pp6t1co14}). \TeXButton{BWE}
{\hfill
\vspace{2ex}
$\blacksquare$}

\TeXButton{Abst}{\vspace{0.8ex}}--- Any two lifts $\hat \phi $ and $\hat
\phi ^{\prime }$ must coincide on the identity component $G_0$, by
uniqueness. However, they may differ on the components $G_0\cdot \kappa $, $%
\kappa \in Ds$, by deck transformations. This gives rise to different
cocycles (in case that ${\cal D}$ is Abelian) $\Gamma $ and $\Gamma ^{\prime
}$. We now show that $\Gamma $ and $\Gamma ^{\prime }$ must differ by a
coboundary:

\subsection{Theorem}

Let the deck transformation group ${\cal D}$ of the covering $p:X\rightarrow
Y$ be Abelian. Given two lifts $\hat \phi $ and $\hat \phi ^{\prime }$ of
the action $\phi :G\times Y\rightarrow Y$ to $X$, with associated cocycles $%
\Gamma $ and $\Gamma ^{\prime }$, there exists a $1$-cochain $\eta \in
C^1\left( G,{\cal D}\right) $ in the ${\cal D}$-valued cohomology on $G$ as
defined in Appendix, section \ref{DeckTransformationValuedCohomOnG}, such
that 
\begin{equation}
\label{pp6t1lift15}\Gamma ^{\prime }=\Gamma +\delta \eta \quad . 
\end{equation}
On the connected component $G_0$, $\eta =e$.

\TeXButton{Beweis}{\raisebox{-1ex}{\it Proof :}
\vspace{1ex}}

Both $\hat \phi $ and $\hat \phi ^{\prime }$ project down to $\phi $ by
(L1), which implies that $\hat \phi _g^{\prime }\hat \phi _g^{-1}$ is a deck
transformation $\eta \left( g\right) $, hence 
\begin{equation}
\label{pp6t1lift151}\hat \phi _g^{\prime }=\eta \left( g\right) \hat \phi
_g\quad . 
\end{equation}
$\eta $ is constant on connected components, hence is trivial on $G_0$. We
have $\hat \phi _g^{\prime }\hat \phi _h^{\prime }=\Gamma ^{\prime }\left(
g,h\right) \widehat{\phi _{gh}^{\prime }}$, and inserting (\ref{pp6t1lift151}%
) gives%
$$
\eta \left( g\right) \cdot \left[ b\left( g\right) \eta \left( h\right)
\right] \cdot \Gamma \left( g,h\right) \cdot \widetilde{\phi _{gh}}=\Gamma
^{\prime }\left( g,h\right) \cdot \eta \left( gh\right) \cdot \widetilde{%
\phi _{gh}}\quad ; 
$$
hence, on using additive notation for the group composition in ${\cal D}$, 
\begin{equation}
\label{pp6t1lift16}\Gamma ^{\prime }\left( g,h\right) =\Gamma \left(
g,h\right) +\,\stackunder{=\left( \delta \eta \right) \left( g,h\right) }{%
\underbrace{b\left( g\right) \eta \left( h\right) -\eta \left( gh\right)
+\eta \left( g\right) }}\quad , 
\end{equation}
using (\ref{pp6t1co14}). \TeXButton{BWE}
{\hfill
\vspace{2ex}
$\blacksquare$}

\TeXButton{Abst}{\vspace{0.8ex}}Thus, all the different lifts $\hat \phi $
of the group action $\phi $ to $X$ give rise to the same cohomology class $%
\left[ \Gamma \right] \in H^2\left( G,{\cal D}\right) $ in the ${\cal D}$%
-valued cohomology on $G$, which also implies, that the possible group
extensions $\tilde G$ of $G$ associated with these lifts are all equivalent 
\cite{Azcarraga}. Thus, up to equivalence, there is only one group extension 
$\tilde G$ of $G$ by ${\cal D}$, and this is basically determined by the
geometry of the covering $p:X\rightarrow Y$.

\section{$\tilde G$-spaces and equivariant covering maps \label
{GSpacesCoveringMaps}}

Although property (L1), $p\hat \phi _g=\phi _gp$, seems to suggest that
every covering map $p$ is a $G$-morphism, this is not true in general, if $G$
is not connected; for in this case, $X$ is not a $G$-space, but only a $%
\tilde G$-space, as explained above. However, we can make $p$ equivariant
with respect to the larger group $\tilde G$: To this end, we first note that
this requires $X$ and $Y$ to be $\tilde G$-spaces. This can be accomplished
by introducing the projection $pr:\tilde G\rightarrow \tilde G/{\cal D}=G$
and observing that $Y$ is trivially a $\tilde G$-space by defining the
action of $\tilde G$ on $Y$ as 
\begin{equation}
\label{pp6t1lift17a}\Phi :\tilde G\times Y\rightarrow Y\quad ,\quad \Phi
\equiv \phi \left( pr\times id_Y\right) \quad . 
\end{equation}
Now we must define a suitable action $\hat \Phi $ of $\tilde G$ on $X$; this
is accomplished by the assignment 
\begin{equation}
\label{pp6t1lift17}\hat \Phi :\tilde G\times X\rightarrow X\quad ,\quad
\left( \left( \gamma ,g\right) ,x\right) \mapsto \hat \Phi \left( \left(
\gamma ,g\right) ,x\right) \equiv \gamma \circ \hat \phi _g\left( x\right)
\quad . 
\end{equation}
Using the group law (\ref{pp6t1lift10}) for the case that ${\cal D}$ is
Abelian we see that 
\begin{equation}
\label{pp6t1lift18}\hat \Phi _{\left( \gamma ,g\right) }\hat \Phi _{\left(
\gamma ^{\prime },g^{\prime }\right) }=\hat \Phi _{\left( \gamma +b\left(
g\right) \gamma ^{\prime }+\Gamma \left( g,g^{\prime }\right) ,gg^{\prime
}\right) }=\hat \Phi _{\left( \gamma ,g\right) \left( \gamma ^{\prime
},g^{\prime }\right) }\quad , 
\end{equation}
i.e. $\hat \Phi $ is indeed a left action. Furthermore, under the projection 
$p$ we have%
$$
p\hat \Phi \left( \left( \gamma ,g\right) ,x\right) =p\hat \phi _g\left(
x\right) =\phi \left( g,p\left( x\right) \right) =\phi \left( pr\times
p\right) \left( \left( \gamma ,g\right) ,x\right) \quad , 
$$
or 
\begin{equation}
\label{pp6t1lift19}p\circ \hat \Phi =\phi \circ \left( pr\times p\right)
=\Phi \circ \left( id_{\tilde G}\times p\right) \quad . 
\end{equation}
But this equation now says that the covering projection $p$ is a $\tilde G$%
-morphism with respect to the group $\tilde G$, where $X$ and $Y$ are now
regarded as $\tilde G$-spaces. Let us summarize:

\subsection{Theorem: Equivariance of the covering map}

Define actions $\Phi ,\hat \Phi $ of the extended group $\tilde G$ on $Y,X$
according to formulas (\ref{pp6t1lift17a}), (\ref{pp6t1lift17}). With this
definition, $X$ and $Y$ become $\tilde G$-spaces, and the covering map $%
p:X\rightarrow Y$ is $\tilde G$-equivariant (a $\tilde G$-morphism).

\TeXButton{Abst}{\vspace{0.8ex}}--- We now discuss a situation in which a
lift $\hat \phi $ as introduced in section \ref{LiftOfGroupActions} always
exists. A glance at formula (\ref{pp6t1lift3}) in section \ref
{ConditionForGroupLift} shows that the lift exists if the fundamental groups
of both $G$ and $X$ are trivial. Now we perform the following construction:
We assume that $G$ is connected and $X$ is simply connected. If $G$ acts on $%
Y$ via $\phi $, then so does the universal simply connected covering group $%
CG$ of $G$; just let $pro:CG\rightarrow G$ be the natural projection (which
is a homomorphism), then $c\phi :CG\times Y\rightarrow Y$, $c\phi \equiv
\phi \left( pro\times id_Y\right) $ defines a smooth left action of $CG$ on $%
Y$. Therefore, in this case there always exists a lift $\widehat{c\phi }%
:CG\times X\rightarrow X$ of $c\phi $ to $X$ satisfying (L1,L2). Since $G$
is connected, formula (\ref{pp6t1lift12a}) in theorem \ref{ExtensionOfG}
tells us that the lift preserves the group law of $CG$, and hence $\widehat{%
c\phi }$ is an action of $CG$ on $X$. Thus,

\subsection{Theorem}

Let $p:X\rightarrow Y$ be a covering, where $X$ is simply connected. Let the
connected Lie group $G$ act on $Y$ via $\phi $, and let $CG$ denote the
universal covering group of $G$ with projection homomorphism $%
pro:CG\rightarrow G$. Then $c\phi \equiv \phi \left( pro\times id_Y\right) $
defines an action of $CG$ on $Y$, and there exists a lift $\widehat{c\phi }%
:CG\times X\rightarrow X$ such that $\widehat{c\phi }$ preserves the group
law on $CG$, 
\begin{equation}
\label{pp6t1lift20}\widehat{c\phi _{gh}}=\widehat{c\phi _g}\circ \widehat{%
c\phi _h}\quad . 
\end{equation}
Thus, the projection map $p$ is a $CG$-morphism of $CG$-spaces, 
\begin{equation}
\label{pp6t1lift21}p\circ \widehat{c\phi _g}=c\phi _g\circ p\quad , 
\end{equation}
for $g\in CG$.

\TeXButton{Abst}{\vspace{0.8ex}}--- A consequence of the developments in
this section is this: Assume that $p:X\rightarrow Y$ is a smooth covering of
manifolds, where $Y$ is a symplectic manifold with symplectic $2$-form $%
\omega $. Since $p$ is a local diffeomorphism, the $2$-form $\Omega \equiv
p^{*}\omega $ is closed and non-degenerate, and hence is a valid symplectic $%
2$-form on $X$. Now assume that a connected Lie group $G$ acts on $Y$ via $%
\Phi :G\times Y\rightarrow Y$ such that all diffeomorphisms $\Phi _g$ are
canonical transformations, $\Phi ^{*}\omega =\omega $. Assume that a lift $%
\tilde \Phi $ to $X$ exists; as $G$ is connected, the group law is then
preserved. Since the lift obeys $p\circ \tilde \Phi _g=\Phi _g\circ p$, it
follows that $\tilde \Phi _g^{*}p^{*}\omega =p^{*}\Phi _g^{*}\omega $, or $%
\tilde \Phi _g^{*}\Omega =\Omega $; hence all diffeomorphisms $\tilde \Phi
_g $ are canonical transformations on the symplectic covering manifold $X$.
In summary,

\subsection{Theorem \label{CanonicalTransOnCovering}}

Let $p:X\rightarrow Y$ be a covering of smooth manifolds, where $Y$ is a
symplectic manifold with symplectic $2$-form $\omega $. Assume the connected
Lie group $G$ acts via $\Phi $ on $Y$ from the left, such that all $\Phi _g$
are canonical transformations with respect to $\omega $. Assume a lift $%
\tilde \Phi $ to $X$ exists. Then all diffeomorphisms $\tilde \Phi _g$ are
canonical transformations with respect to the symplectic $2$-form $\Omega
\equiv p^{*}\omega $ on $X$.

\TeXButton{Abst}{\vspace{0.8ex}}--- Finally, we wish to study the following
problem: Given a non-simply connected manifold $Y$, thought of as the
''configuration space'' of a dynamical system (or rather, a class of
dynamical systems), and the left action $\phi :G\times Y\rightarrow Y$ of a
(Lie) group $G$ on $Y$; if a lift $\hat \phi $ exists, we can extend it to a
map $^{*}\hat \phi :G\times T^{*}X\rightarrow T^{*}X$ on the cotangent
bundle of $X$ using formula (\ref{pp6t1cot1}). On the other hand, from the
discussion in section \ref{PStarIsACovering} we know that $T^{*}X$ is itself
a covering space of $T^{*}Y$, and $^{*}p$ as defined in equation (\ref
{pp6t1pstarisacovering}) is the covering map. Hence we can extend the action 
$\phi $ first to an action $^{*}\phi $ on the cotangent bundle $T^{*}Y$, and
then ask whether a lift $\widehat{^{*}\phi }$ of $^{*}\phi $ to $T^{*}X$
exists, and if yes, whether it coincides with the extension $^{*}\hat \phi $
of the lift $\hat \phi $ of $\phi $. The next theorem gives the answer for
connected $G$:

\subsection{Theorem \label{LiftsAndExtensions}}

Let $G$ be a connected Lie group; let $p:X\rightarrow Y$ be a covering,
where $X$ is connected. Let $\phi :G\times Y\rightarrow Y$ be a smooth
action of $G$ on $Y$, which possesses a lift $\hat \phi :G\times
X\rightarrow X$ satisfying (L1, L2). Then $\left( 1\right) $ a lift $%
\widehat{^{*}\phi }:G\times T^{*}X\rightarrow T^{*}X$ of the extension $%
^{*}\phi :G\times T^{*}Y\rightarrow T^{*}Y$ of the action $\phi $ to the
cotangent bundle $T^{*}X$ of $X$ exists; and $\left( 2\right) $ this lift
coincides with the uniquely determined extension $^{*}\widehat{\phi }%
:G\times T^{*}X\rightarrow T^{*}X$ of the lift $\hat \phi :G\times
X\rightarrow X$ of $\phi $ to the cotangent bundle $T^{*}X$ of $X$; i.e. 
\begin{equation}
\label{pp6t1lift22}\widehat{^{*}\phi }=\,^{*}\widehat{\phi }\quad , 
\end{equation}
yielding a commutative diagram.

\TeXButton{Beweis}{\raisebox{-1ex}{\it Proof :}
\vspace{1ex}}

Let $\hat \phi $ be the lift of $\phi $ satisfying (L1,L2). Its extension to
the cotangent bundle $T^{*}X$ takes the form $^{*}\widehat{\phi }=\left(
\hat \phi ,\hat \phi ^{-1*}\right) $. This map satisfies 
\begin{equation}
\label{pp6t1lift23a}^{*}\widehat{\phi _e}=id_{T^{*}X}\quad , 
\end{equation}
and under the extended projection $^{*}p$ it behaves as 
\begin{equation}
\label{pp6t1lift23}^{*}p\circ \,^{*}\hat \phi =\,^{*}\phi \circ \,\left(
id_G\times \,^{*}p\right) \quad . 
\end{equation}
But formulas (\ref{pp6t1lift23a},\ref{pp6t1lift23}) are precisely the
conditions (L1, L2) for a lift $\widehat{^{*}\phi }$ of the extended action $%
^{*}\phi $ under the covering $^{*}p:T^{*}X\rightarrow T^{*}Y$. This means
that $\left( 1\right) $ $^{*}\hat \phi $ is a lift of $^{*}\phi $; and,
since $G$ and $X$, and hence $G\times T^{*}X$, are connected, this lift is
unique as a consequence of the ''Unique Lifting Theorem'' \ref
{UniqueLiftingTheorem}, and hence coincides with $\widehat{^{*}\phi }$. 
\TeXButton{BWE}{\hfill
\vspace{2ex}
$\blacksquare$}

\section{Symplectic $G$-actions and moment maps \label
{SymplecticGActionsMoMaps}}

In the following sections we generalize the usual definition of a moment map
to a construction we call local moment map. The standard definitions found
in the literature generally involve that a moment map is a {\bf globally}
defined function from a symplectic manifold $M$ into the coalgebra of the
Lie algebra of a Lie group $G$ which acts on $M$ symplectically (e.g. \cite
{AbrahamMarsden}). Other authors (e.g. \cite{GuillStern}, \cite{Woodhouse})
give even more restrictions by introducing moment maps only together with
the condition that the first and second Chevalley-Eilenberg cohomology
groups of the Lie group $G$ are trivial, or equivalently, that the
associated Lie algebra cohomoloy groups $H_0^1\left( \hat g,\TeXButton{R}
{\mathbb{R}}\right) $ and $H_0^2\left( \hat g,\TeXButton{R}{\mathbb{R}}%
\right) $ are trivial. In this work we make no assumptions about cohomologies on the group $G$,
nor do we assume that moment maps exist globally; on the contrary, it is the
purpose of this work to generalize moment maps to situations where the
underlying symplectic manifold is non-simply connected, and hence in general
does not admit a global moment map.

Let $M$ be a symplectic manifold with symplectic form $\omega $. Let $G$ be
a Lie group, let $\Phi :G\times M\rightarrow M$ be a smooth symplectic left
action of $G$ on $M$, i.e. $\Phi \left( g,x\right) =\Phi _g\left( x\right) $%
, with $\Phi _{gg^{\prime }}=\Phi _g\Phi _{g^{\prime }}$, $\Phi _e=id_M$,
and $\Phi _g^{*}\omega =\omega $ for all $g$. Let $\hat g$ be the Lie
algebra of $G$, and $g^{*}$ denote the coalgebra. If $A\in \hat g$, then the
vector field induced by $A$ on $M$ is denoted $\frac{\partial \Phi }{%
\partial G}A\equiv \tilde A$. Since $\Phi $ preserves the symplectic form,
the Lie derivative of $\omega $ with respect to $\tilde A$ vanishes, hence $%
\tilde A$ is a locally Hamiltonian vector field according to (\ref{pp6t1fo2}%
). From this it follows that the $1$-form $\frac{\partial \Phi }{\partial G}A%
\TeXButton{i}{\intmul}\omega $ is closed.

\section{Global moment maps \label{GlobalMomentMaps}}

First we assume that $M$ is simply connected. By a standard argument, an $%
\TeXButton{R}{\mathbb{R}}$-linear map $f$ from a vector space $\hat g$ to
the space of smooth closed $1$-forms $Z_{deRham}^1\left( M\right) $ on $M$
can always be lifted to an $\TeXButton{R}{\mathbb{R}}$-linear map $h:\hat
g\rightarrow {\cal F}\left( M\right) $, $A\mapsto h_A$ such that $%
dh_A=f\left( A\right) $: For, since $M$ is simply connected, every closed $1$%
-form $f\left( A\right) $ has a potential $h_A$ with $dh_A=f\left( A\right) $%
; the assignment $\left( A,x\right) \mapsto h_A\left( x\right) $ can be
assumed to be smooth in $x$, but need not be smooth in $A$. Now choose an
arbitrary fixed point $x_0\in M$, and replace $h_A$ by $h_A-h_A\left(
x_0\right) $; then $A\mapsto h_A\left( x\right) -h_A\left( x_0\right) $ is
linear in $A$.

In particular, the map $\hat g\ni A\mapsto \frac{\partial \Phi }{\partial G}A%
\TeXButton{i}{\intmul}\omega \in Z^1\left( M\right) $ can be lifted to an $%
\TeXButton{R}{\mathbb{R}}$-linear map $h:\hat g\rightarrow {\cal F}\left(
M\right) $, $A\mapsto -h_A$, with 
\begin{equation}
\label{pp6t1sym1}\frac{\partial \Phi }{\partial G}A\TeXButton{i}{\intmul}%
\omega +dh_A=0\quad . 
\end{equation}
Since $A\mapsto h_A\left( x\right) $ is linear for every fixed $x\in M$, $h$
defines a map $J:M\rightarrow g^{*}$, $\left\langle J\left( x\right)
,A\right\rangle \equiv h_A\left( x\right) $. $J$ is called a {\it moment map}
associated with the action $\Phi $. From its definition via $h$ we see that $%
J$ is determined up to addition $J\mapsto J+L$ of an $M$-constant, $%
\TeXButton{R}{\mathbb{R}}$-linear map $L:\hat g\rightarrow \TeXButton{R}
{\mathbb{R}}$, with $dL=0$.

\TeXButton{Abst}{\vspace{0.8ex}}$h$ as defined above is a homomorphisms of
vector spaces by linearity; in general it is not a homomorphism of Lie
algebras, however. Rather, the algebra of Poisson brackets can provide a
central extension of the Lie algebra of the Hamiltonian vector fields \cite
{GuillStern}.

\section{Local moment maps \label{LocalMomentMaps}}

Now we assume that $Y$ is connected, but not simply connected. If $y$ is a
base point of $Y$, set ${\cal D}\equiv \pi _1\left( Y,y\right) $. Let ${\cal %
V}=\left\{ V_a\mid a\in A\right\} $ be a countable simply connected open
cover of $Y$ as introduced in section \ref{MultiValFunc}. For every $A\in
\hat g$, $\frac{\partial \Phi }{\partial G}A\TeXButton{i}{\intmul}\omega $
is a closed $1$-form on $Y$; hence a multi-valued potential function $\left(
h_{A,a,d}\right) $, $d\in {\cal D}$, exists for every \u Cech cocycle $%
\left( g_{ab}\right) $ associated with a simply connected cover of $Y$,
according to theorem \ref{MultiValuedPotential}. However, here we can no
longer be sure whether all $h_{A,a,d}$ can be made linear in $A$
simultaneously, without spoiling the glueing conditions $h_{A,a,d}=h_{A,b,d%
\cdot g_{ab}}$. We therefore have to formulate the problem in terms of an
appropriate covering space $X$, which we take, as in the proof of theorem 
\ref{MultiValuedPotential}, to be the identification space $i_a:{\cal D}%
\times V_a\rightarrow X\equiv \bigsqcup\limits_{a\in A}{\cal D}\times
V_a/\sim $, where $\sim $ relates elements $\left( d,y\right) $ and $\left(
d^{\prime },y^{\prime }\right) =\left( d\cdot g_{ab},y\right) $ on ${\cal D}%
\times V_a$ and ${\cal D}\times V_b$ that are identified as $i_a\left(
d,y\right) =i_b\left( d^{\prime },y^{\prime }\right) $. Then $X$ is a
universal covering space of $Y$, such that the projection $p:X\rightarrow Y$
is a local diffeomorphism. $Y$ is the space of orbits under the action of $%
{\cal D}$ on $X$. If $\omega $ denotes the symplectic $2$-form on $Y$, $%
\Omega \equiv p^{*}\omega $ is a symplectic $2$-form on $X$.

We must ask whether $\Phi $ can be lifted to $X$, i.e. whether there exists
a map $\hat \Phi :G\times X\rightarrow X$ satisfying $p\circ \hat \Phi =\hat
\Phi \circ \left( id_G\times p\right) $, and $\hat \Phi _e=id_X$.
Furthermore, one has to examine whether the group law is preserved by the
lift, i.e. whether $\widehat{\Phi _{gh}}=\hat \Phi _g\hat \Phi _h$. The
question of existence of a lift is examined in theorem \ref
{ConditionForGroupLift}; in theorem \ref{ExtensionOfG} it is proven that,
for connected $G$, the lift preserves the group law of $G$. In theorem \ref
{CanonicalTransOnCovering} it is proven that, for connected $G$, the
diffeomorphisms $\hat \Phi _g$ are canonical transformations with respect to 
$\Omega $, hence $\hat \Phi $ is a symplectic action. In this case we have a
left action $\hat \Phi $ of $G$ on the simply connected manifold $X$, so
that the results from section \ref{GlobalMomentMaps} can be applied: There
exists a lift of the $\TeXButton{R}{\mathbb{R}}$-linear map $\hat g\ni
A\mapsto \frac{\partial \hat \Phi }{\partial G}A\TeXButton{i}{\intmul}\Omega
\in Z_{deRham}^1\left( X\right) $ to an $\TeXButton{R}{\mathbb{R}}$-linear
map $\hat h:\hat g\rightarrow {\cal F}\left( X\right) $, $A\mapsto -\hat h_A$%
, with 
\begin{equation}
\label{pp6t1LocMom1}\frac{\partial \hat \Phi }{\partial G}A\TeXButton{i}
{\intmul}\Omega +d\hat h_A=0\quad , 
\end{equation}
and there exists a global moment map $\hat J:X\rightarrow g^{*}$, $%
\left\langle \hat J\left( x\right) ,A\right\rangle \equiv \hat h_A\left(
x\right) $. $\hat J$ is determined up to addition $\hat J\mapsto \hat J+L$
of an $X$-constant, $\TeXButton{R}{\mathbb{R}}$-linear map $L:\hat
g\rightarrow \TeXButton{R}{\mathbb{R}}$, with $dL=0$. Now we can set $%
h_{a,d}\left( y\right) \equiv \hat h\circ i_a\left( d,y\right) $, and $%
J_{a,d}\left( y\right) \equiv \hat J\circ i_a\left( d,y\right) $ for $y\in
V_a$, $d\in {\cal D}$. Then 
\begin{equation}
\label{pp6t1LocMom2}\frac{\partial \Phi }{\partial G}A\TeXButton{i}{\intmul}%
\omega +d\left\langle J_{a,d},A\right\rangle =0 
\end{equation}
on $V_a$, and for all $d\in {\cal D}$. The glueing condition is easily
derived as%
$$
J_{a,d}\left( y\right) =\hat J\circ i_a\left( d,y\right) =\left( \hat J\circ
i_b\right) \circ \left( i_b^{-1}\circ i_a\right) \left( d,y\right) = 
$$
\begin{equation}
\label{pp6t1LocMom3}=\left( \hat J\circ i_b\right) \left( d\cdot
g_{ab},y\right) =J_{b,d\cdot g_{ab}}\left( y\right) \quad . 
\end{equation}

--- Now we examine to which extent a collection $\left( J_{a,d}\right) $ is
determined by the action $\Phi $ and a cocycle: Let $\left( g_{ab}\right) $, 
$\left( g_{ab}^{\prime }\right) $ be cocycles such that the associated
homomorphisms $\rho $, $\rho ^{\prime }$ are inner automorphisms of ${\cal D}
$, hence give rise to simply connected coverings; and let $\left(
J_{a,d}\right) $, $\left( J_{a,d}^{\prime }\right) $ be collections
satisfying relations (\ref{pp6t1LocMom2},\ref{pp6t1LocMom3}), respectively.
Then $\left( g_{ab}\right) $ and $\left( g_{ab}^{\prime }\right) $ are
cohomologous. As above, construct smooth simply connected covering manifolds 
$p:X\rightarrow Y$, $q:Z\rightarrow Y$ as identification spaces; the
trivializations $\left( i_a\right) $ with respect to $X$ identify $i_a\left(
d,y\right) =i_b\left( d\cdot g_{ab},y\right) $, and the trivializations $%
\left( j_a\right) $ with respect to $Z$ identify $j_a\left( d,y\right)
=j_b\left( d\cdot g_{ab}^{\prime },y\right) $. A lift $\tilde \Phi $ of $%
\Phi $ to $Z$ exists precisely when a lift $\hat \Phi $ of $\Phi $ to $X$
exists. The glueing conditions (\ref{pp6t1LocMom3}) for $\left(
J_{a,d}\right) $, $\left( J_{a,d}^{\prime }\right) $ guarantee that there
exist smooth functions $\hat h_A$, $\tilde h_A$ on $X$, $Z$, obeying 
\begin{equation}
\label{pp6t1LocMom70}\hat h_A\circ i_{a,d}=\left\langle
J_{a,d},A\right\rangle \text{\quad and\quad }\tilde h_A\circ
j_{a,d}=\left\langle J_{a,d}^{\prime },A\right\rangle \quad . 
\end{equation}
These functions define global moment maps $\left\langle \tilde
J,A\right\rangle =\tilde h_A$ on $Z$ and $\left\langle \hat J,A\right\rangle
=\hat h_A$ on $X$. $\tilde J$ satisfies the analogue of (\ref{pp6t1LocMom1}%
), 
\begin{equation}
\label{pp6t1LocMom7}\frac{\partial \tilde \Phi }{\partial G}A\TeXButton{i}
{\intmul}\Omega ^{\prime }+d\tilde h_A=0 
\end{equation}
on $Z$, where $\Omega ^{\prime }\equiv q^{*}\omega $. Furthermore, $Z$ and $%
X $ are ${\cal D}$-isomorphic, the isomorphism being effected by a
diffeomorphism $\phi :Z\rightarrow X$, where $\phi $ preserves fibres, $%
p\circ \phi =q$, and is ${\cal D}$-equivariant. We examine the relation
between the lifts $\hat \Phi $ and $\tilde \Phi $: Define a map $\psi
:G\times Z\rightarrow Z$, 
\begin{equation}
\label{pp6t1LocMom4}\left( g,z\right) \mapsto \psi \left( g,z\right) \equiv
\phi ^{-1}\circ \hat \Phi \left( g,\phi \left( z\right) \right) \quad . 
\end{equation}
A calculation shows that $\psi $ satisfies $q\circ \psi =\Phi \circ \left(
id_G\times q\right) $, and $\psi _e=id_Z$. But, as discussed in section \ref
{PreservationOfTheGroupLaw}, there exists precisely one function $G\times
Z\rightarrow Z$ with these two properties, and this is just the lift $\tilde
\Phi $! Hence we deduce that $\psi =\tilde \Phi $, or 
\begin{equation}
\label{pp6t1LocMom5}\phi \circ \tilde \Phi =\hat \Phi \circ \left(
id_G\times \phi \right) \quad , 
\end{equation}
which gives a commutative diagram. It follows that 
\begin{equation}
\label{pp6t1LocMom6}\frac{\partial \hat \Phi }{\partial G}=\phi _{*}\frac{%
\partial \tilde \Phi }{\partial G}\quad . 
\end{equation}

The symplectic $2$-forms on $X$ and $Z$ are $\Omega \equiv p^{*}\omega $ and 
$\Omega ^{\prime }\equiv q^{*}\omega $, where $\omega $ is the symplectic $2$%
-form on $Y$. From $p\phi =q$ we infer that $\Omega ^{\prime }=\phi
^{*}\Omega $, hence $\phi $ is a symplectomorphism. From theorem \ref
{CanonicalTransOnCovering} we know that $\hat \Phi $, $\tilde \Phi $ are
symplectic actions with respect to $\Omega $, $\Omega ^{\prime }$.

If we insert (\ref{pp6t1LocMom6}) into (\ref{pp6t1LocMom1}) we obtain 
\begin{equation}
\label{pp6t1LocMom8}\left[ \phi _{*}\frac{\partial \tilde \Phi }{\partial G}%
A\right] \TeXButton{i}{\intmul}\Omega +d\hat h_A=0\quad ; 
\end{equation}
using the relation $\Omega ^{\prime }=\phi ^{*}\Omega $ and (\ref
{pp6t1LocMom7}) we deduce $\left[ \phi _{*}\frac{\partial \tilde \Phi }{%
\partial G}A\right] \TeXButton{i}{\intmul}\Omega =\phi ^{-1*}\left[ \frac{%
\partial \tilde \Phi }{\partial G}A\TeXButton{i}{\intmul}\Omega ^{\prime
}\right] $, which, together with (\ref{pp6t1LocMom8}), yields $d\left\langle
\tilde J,A\right\rangle =\phi ^{*}d\left\langle \hat J,A\right\rangle $, or 
\begin{equation}
\label{pp6t1LocMom9}\tilde J=\hat J\circ \phi +L\quad , 
\end{equation}
where $L$ is a $Z$-constant linear map $\hat g\rightarrow \TeXButton{R}
{\mathbb{R}}$. Using the trivializations $\left( i_a\right) $, $\left(
j_a\right) $, (\ref{pp6t1LocMom70}) and (\ref{pp6t1LocMom9}) we have%
$$
J_{a,d}^{\prime }\left( y\right) =\tilde J\circ j_{a,d}=\left[ \hat J\circ
\phi +L\right] \circ j_{a,d}=\left( \hat J\circ i_a\right) \circ \left(
i_a^{-1}\circ \phi \circ j_a\right) \left( d,y\right) +L\quad . 
$$
As in the proof of theorem \ref{MultiValuedPotential}, $\left( i_a^{-1}\circ
\phi \circ j_a\right) \left( d,y\right) =\left( d\cdot k_a,y\right) $ for a $%
0$-\u Cech cochain $\left( k_a\right) $ which is determined by the cocycles $%
\left( g_{ab}\right) $, $\left( g_{ab}^{\prime }\right) $ up to its value in
a coset of the center of ${\cal D}$ in ${\cal D}$. Then the last equation
gives%
$$
J_{a,d}^{\prime }=J_{a,d\cdot k_a}+L\quad . 
$$
In summary, we have proven:

\subsection{Theorem and Definition: Local moment map \label
{TheoremLocalMomentMap}}

Let $Y$ be a connected, but not simply connected symplectic manifold with
base point $y$, with ${\cal D}\equiv \pi _1\left( Y,y\right) $, and $\omega $
is the symplectic $2$-form on $Y$. Let $\Phi :G\times Y\rightarrow Y$ be a
symplectic left action of a connected Lie group $G$ on $Y$ with respect to $%
\omega $. Let ${\cal V}=\left\{ V_a\mid a\in A\right\} $ be a simply
connected path-connected open cover of $Y$. Then

\begin{description}
\item[(A)]  \quad for every ${\cal D}$-valued $1$-\u Cech-cocycle $\left(
g_{ab}\right) $, $a,b\in A$, on ${\cal V}$, whose associated homomorphism $%
\rho :{\cal D}\rightarrow {\cal D}$ is an {\it inner} automorphism of ${\cal %
D}$, there exists a collection $\left( J_{a,d}\right) $ of coalgebra-valued
functions $J_{a,d}:V_a\rightarrow g^{*}$ for $a\in A$, $d\in {\cal D}$, such
that

\begin{enumerate}
\item  
\begin{equation}
\label{pp6t1LocMom10}\frac{\partial \Phi }{\partial G}A\TeXButton{i}{\intmul}%
\omega +d\left\langle J_{a,d},A\right\rangle =0 
\end{equation}
on $V_a$, and for all $d\in {\cal D}$;

\item  \quad let $\lambda $ be a loop at $y$ with $\left[ \lambda \right]
=d\in \pi _1\left( Y,y\right) \simeq {\cal D}$. Then 
\begin{equation}
\label{pp6t1SheetsOfMoment}\left\langle J_{a,d},A\right\rangle =\left\langle
J_{a,e},A\right\rangle -\int\limits_\lambda \frac{\partial \Phi }{\partial G}%
A\TeXButton{i}{\intmul}\omega 
\end{equation}
for all $A\in \hat g$, where $e$ is the identity in ${\cal D}$.

\item  \quad the $J_{a,d}$ satisfy a {\it glueing condition}, expressed by 
\begin{equation}
\label{pp6t1LocMom11}J_{a,d}=J_{b,d\cdot g_{ab}} 
\end{equation}
on $V_a\cap V_b\neq \emptyset $.
\end{enumerate}

\item[(B)]  \quad Let $\left( g_{ab}^{\prime }\right) $ be another cocycle
giving rise to a simply connected cover of $Y$, and let $\left(
J_{a,d}^{\prime }\right) $ be another collection of functions on ${\cal V}$
satisfying properties (A1--A3) with respect to $\left( g_{ab}^{\prime
}\right) $ and the action $\Phi $. Then there exists a $Y$-constant linear
map $L:\hat g\rightarrow \TeXButton{R}{\mathbb{R}}$ (i.e. $dL=0$) and a $%
{\cal D}$-valued $0$-\u Cech cochain $\left( k_a:V_a\rightarrow {\cal D}%
\right) $ on ${\cal V}$ such that 
\begin{equation}
\label{pp6t1LocMom12}J_{a,d}^{\prime }=J_{a,d\cdot k_a}+L 
\end{equation}
for all $a\in A$, $d\in {\cal D}$. The $0$-cochain $\left( k_a\right) $ is
determined by the cocycles $\left( g_{ab}\right) $ and $\left(
g_{ab}^{\prime }\right) $ as expressed in theorem \ref{MultiValuedPotential}.

\item[(C)]  \quad {\bf Definition:\quad }A collection $\left(
g_{ab};J_{a,d}\right) $ satisfying properties (A1--A3) will be called a {\it %
local moment map} for the action $\Phi $ on the symplectic manifold $\left(
Y,\omega \right) $.
\end{description}

\section{Equivariance of moment maps \label{EquivarianceOfMomentMaps}}

Usually, moment maps are introduced in a more restricted context. For
example, conditions are imposed from the start so as to guarantee the
existence of a uniquely determined single-valued globally defined moment
map. Furthermore, it is often assumed that the first and second
Chevalley-Eilenberg cohomology groups of the group $G$ vanish, which then
provides a sufficient condition for the moment map to transform as a $G$%
-morphism \cite{Woodhouse},\cite{GuillStern}. Our approach will be slightly
more general. The first generalization has been made above, allowing for
moment maps to be only locally defined. The second
one is, that we do not want to enforce the moment maps to behave as strict $%
G $-morphisms; rather, the deviation from transforming equivariantly is
determined by a cocycle in a certain cohomology on $\hat g$, which in turn
gives rise to a central extension of the original Lie algebra $\hat g$,
which is interesting in its own right and also physically relevant.

Let the conditions of theorem \ref{TheoremLocalMomentMap} be given.
Reconstruct a simply connected cover $p:X\rightarrow Y$ from ${\cal V}$ and $%
\left( g_{ab}\right) $ as in section \ref{LocalMomentMaps}, together with
trivializations $\left( i_a\right) $ such that $i_b^{-1}\circ i_a\left(
d,y\right) =\left( d\cdot g_{ab},y\right) $. Assume a lift $\hat \Phi _g$
exists. Assume that $p\circ \hat \Phi _g\left( x\right) \in V_b$, where $%
x=i_a\left( d,y\right) $; then there exists a unique $d^{\prime }\in {\cal D}
$ with $i_b^{-1}\circ \hat \Phi _g\left( x\right) =\left( d^{\prime },\Phi
_g\left( y\right) \right) $, with $\Phi _g\left( y\right) \in V_b$. Here $%
d^{\prime }$ is a function of $d$, $g$, and $y$; its structure can be
understood from the following consideration: Let $\lambda $ be a path in $G$
connecting $e$ with $g$. Then $t\mapsto \Phi \left( \lambda \left( t\right)
,y\right) $ is a path in $Y$ connecting $y$ with $\Phi _g\left( y\right) $,
which has a unique lift to $i_a\left( d,y\right) $, whose endpoint is just $%
\hat \Phi _g\circ i_a\left( d,y\right) $, by the definition of the lift $%
\hat \Phi $. The condition $\Phi _{y\#}\pi _1\left( G,e\right) \subset \pi
_1\left( Y,y\right) $ guarantees that this construction is independent of
the path $\lambda $. The interval $\left[ 0,1\right] $ can be partitioned
into subintervals $\left[ a_{i-1},a_i\right] $ so that $\Phi \left( \lambda
\left( t\right) ,y\right) \in V_{a_i}$ for $t\in \left[ a_{i-1},a_i\right] $%
, and $\lambda \left( t_n\right) =g$. It follows that 
\begin{equation}
\label{pp6t1equi2}i_{a_n}^{-1}\circ \hat \Phi _g\left( x\right) =\left(
d\cdot g_{a_0a_1}\cdots g_{a_{n-1}a_n},\Phi _g\left( y\right) \right) \quad
, 
\end{equation}
and hence $d^{\prime }=d\cdot g_{a_0a_1}\cdots g_{a_{n-1}a_n}\equiv d\cdot
\psi _{a_n}\left( g,y\right) $. Since ${\cal D}$ is discrete, $\psi _{a_n}$
is locally constant. Altogether we have shown that if $p\circ \hat \Phi
_g\left( x\right) \in V_b$, then 
\begin{equation}
\label{pp6t1equi3}i_b^{-1}\circ \hat \Phi _g\left( x\right) =\left( d\cdot
\psi _b,\Phi _g\left( y\right) \right) \quad . 
\end{equation}

\TeXButton{Abst}{\vspace{0.8ex}}--- Next we recall without proof the $G$%
-transformation behaviour for global moment maps (see \cite{Azcarraga}):

\subsection{Theorem: $G$-transformation of global moment maps \label
{GTransGlobalMomentMap}}

Let $\left( M,\omega \right) $ be a simply connected symplectic manifold,
let $\Phi :G\times M\rightarrow M$ be a symplectic left action of the Lie
group $G$ on $M$. Let $J$ be a global moment map for the action $\Phi $.
Then 
\begin{equation}
\label{pp6t1equi4}J\circ \Phi _g=Ad^{*}\left( g\right) \cdot J+\alpha \left(
g\right) \quad , 
\end{equation}
where $\alpha :G\rightarrow g^{*}$ is a $1$-cocycle in the $g^{*}$-valued
cohomology on $G$ as defined in section \ref{gStarOnGroup}, i.e. $\alpha \in
Z^1\left( G,g^{*}\right) $. This means that 
\begin{equation}
\label{pp6t1equi5}\left( \delta \alpha \right) \left( g,h\right)
=Ad^{*}\left( g\right) \cdot \alpha \left( h\right) -\alpha \left( gh\right)
+\alpha \left( g\right) =0 
\end{equation}
for all $g,h\in G$. Thus, (\ref{pp6t1equi4}) says that $J$ transforms
equivariantly under $G$ up to a cocycle in the $g^{*}$-valued cohomology.

\TeXButton{Abst}{\vspace{0.8ex}}--- Now we generalize this result to local
moment maps. We prove:

\subsection{Theorem: $G$-transformation behaviour of local moment maps \label
{GTransLocalMomentMap}}

Let $Y$ be a connected symplectic manifold, with ${\cal D}\equiv \pi
_1\left( Y,y\right) $. Let $\Phi :G\times Y\rightarrow Y$ be a symplectic
left action of a connected Lie group $G$ on $Y$. Let ${\cal V}=\left\{
V_a\mid a\in A\right\} $ be a simply connected open cover of $Y$. Let $%
\left( g_{ab}\right) $, $a,b\in A$, be a ${\cal D}$-valued $1$-\u
Cech-cocycle on ${\cal V}$ describing a simply connected ${\cal D}$-covering
space of $Y$, and let the collection $\left( J_{a,d};g_{ab}\right) $ be the
associated local moment map with respect to the action $\Phi $.

Let $y\in Y$, and assume that $p\circ \Phi _g\left( y\right) \in V_b$. Then 
\begin{equation}
\label{pp6t1GTrans1}J_{b,d\cdot \psi _b}\circ \Phi _g=Ad^{*}\left( g\right)
\cdot J_{a,d}\;+\;\alpha \left( g\right) \quad , 
\end{equation}
where $\alpha :G\rightarrow g^{*}$ is a $1$-cocycle in the $g^{*}$-valued
cohomology on $G$ as defined in appendix, section \ref{gStarOnGroup}, and $%
\psi _g$ is defined in formula (\ref{pp6t1equi3}).

\TeXButton{Beweis}{\raisebox{-1ex}{\it Proof :}
\vspace{1ex}}

Let the conditions of theorem \ref{TheoremLocalMomentMap} be given, and
reconstruct a simply connected cover $p:X\rightarrow Y$ from ${\cal V}$ and $%
\left( g_{ab}\right) $ as in section \ref{LocalMomentMaps}, together with
trivializations $\left( i_a\right) $ such that $i_b^{-1}\circ i_a\left(
d,y\right) =\left( d\cdot g_{ab},y\right) $. Assume a lift $\hat \Phi _g$
exists. By $\hat J$ we denote the global moment map on $X$ with respect to $%
\hat \Phi $. This satisfies 
\begin{equation}
\label{pp6t1GTrans2}\hat J\circ \hat \Phi _g=Ad^{*}\left( g\right) \cdot
\hat J+\alpha \left( g\right) \quad , 
\end{equation}
according to (\ref{pp6t1equi4}). From (\ref{pp6t1equi3}) it follows that 
\begin{equation}
\label{pp6t1GTrans3}i_b^{-1}\circ \hat \Phi _g\circ i_a\left( d,y\right)
=\left( d\cdot \psi _b,\Phi _g\left( y\right) \right) \quad . 
\end{equation}
Using $J_{a,d}=\hat J\circ i_{a,d}$ in (\ref{pp6t1GTrans2}), (\ref
{pp6t1GTrans3}) implies%
$$
\left[ \hat J\circ i_b\right] \circ \left[ i_b^{-1}\circ \hat \Phi _g\circ
i_a\right] \left( d,y\right) =J_{b,d\cdot \psi _b}\left( y\right) 
$$
for the LHS, and hence the result (\ref{pp6t1GTrans1}). \TeXButton{BWE}
{\hfill
\vspace{2ex}
$\blacksquare$}

\section{Non-simply connected coverings \label{NonSimplyConnected}}

In this section we study the relation between local moment maps on
symplectic manifolds $Z$, $Y$ where $q:Z\rightarrow Y$ is a covering of
manifolds, but $Z$ is not necessarily simply connected:

Let $\zeta \in Z$, $\eta \in Y$ be base points with $q\left( \zeta \right)
=\eta $; let ${\cal D}\equiv \pi _1\left( Y,\eta \right) $, and ${\cal H}%
\equiv \pi _1\left( Z,\zeta \right) $. Let $X$ be a universal covering
manifold $p:X\rightarrow Y$ of $Y$ such that $Y$ is the orbit space $Y=X/%
{\cal D}$. Then $X$ is also a universal cover of $Z$, and $Z$ is isomorphic
to the orbit space $X/{\cal H}$; for the sake of simplicity we ignore this
isomorphism and identify $Z=X/{\cal H}$. There is a covering projection $%
r:X\rightarrow Z$, taking the base point $\xi \in X$ to $r\left( \xi \right)
=\zeta $, and $p=q\circ r$. Since $q_{\#}\pi _1\left( Z,\zeta \right) $ is
an injective image of ${\cal H}$ in ${\cal D}$, we identify ${\cal H}$ with
its image under $q_{\#}$, and thus can regard ${\cal H}$ as a subgroup of $%
{\cal D}$.

If $\omega $ is a symplectic form on $Y$, then $\Omega \equiv q^{*}\omega $
and $\hat \Omega \equiv p^{*}\omega $ are the natural symplectic forms on $Z$
and $X$, respectively, and $\hat \Omega =r^{*}\Omega $.

Let the Lie group $G$ act on $Y$ from the left via $\Phi :G\times
Y\rightarrow Y$. We assume that the lift $\hat \Phi $ of $\Phi $ to $X$
exists, which is true if and only if $\Phi _{\eta \#}\pi _1\left( G,e\right)
=\left\{ e\right\} $. It is easy to show that in this case the lift $\tilde
\Phi $ of $\Phi $ to $Z$ exists; and furthermore, that the lift of $\tilde
\Phi $ to $X$ coincides with $\hat \Phi $.

We now introduce trivializations for $p:X\rightarrow Y$, and subsequently,
construct preferred trivializations of the coverings $r:X\rightarrow Z$ and $%
q:Z\rightarrow Y$ based on this. Firstly, let ${\cal V}$ be a simply
connected open cover of $Y$ as above, and let $\left( i_a:{\cal D}\times
V_a\rightarrow X\right) $ be a trivialization of the covering $%
p:X\rightarrow Y$. Given an element $d\in {\cal D}$, let $\left[ d\right] $
denote the coset $\left[ d\right] \equiv {\cal H}\cdot d$, where ${\cal H}%
\subset {\cal D}$ is regarded as a subgroup of ${\cal D}$, as above. For $%
V_a\in {\cal V}$ and $d\in {\cal D}$, define sets $U_{a,\left[ d\right]
}\equiv r\circ i_a\left( \left\{ d\right\} \times V_a\right) \subset Z$. The
sets $U_{a,\left[ d\right] }$ are open and simply connected by construction,
and their totality 
\begin{equation}
\label{pp6t1nonsim1}{\cal U}\equiv \left\{ U_{a,\left[ d\right] }\mid a\in
A\,;\;\left[ d\right] \in {\cal D/H}\right\} 
\end{equation}
covers $Z$, since the sets $i_a\left( d,V_a\right) $ cover $X$. The $%
U_{a,\left[ d\right] }$ are just the connected components of the inverse
image $q^{-1}\left( V_a\right) \subset Z$, hence we have $q\left(
U_{a,\left[ d\right] }\right) =V_a$ for all $\left[ d\right] \in {\cal D}/%
{\cal H}$, and ${\cal U}$ is a simply connected open (countable) cover of $Z$%
. We define a trivialization $\left( k_a:{\cal D}/{\cal H}\times
V_a\rightarrow Z\right) $ of the covering $q:Z\rightarrow Y$ by $k_a\left(
\left[ d\right] ,y\right) \equiv r\circ i_a\left( d,y\right) $. Since $r$
maps points $x\in X$ into orbits $r\left( x\right) =\hat \Phi \left( {\cal H}%
,x\right) $, this definition is independent of the representative $d$ of $%
\left[ d\right] $. Now we can construct trivializations of $r:X\rightarrow Z$
based on the cover ${\cal U}$ of $Z$; in particular, by specifying
representatives $d_0$ of the various cosets $\left[ d_0\right] $, we see
that there exists a trivialization $\left( j_{a,\left[ d_0\right] }:{\cal H}%
\times U_{a,\left[ d_0\right] }\rightarrow X\right) $ such that 
\begin{equation}
\label{pp6t1nonsim2}j_{a,\left[ d_0\right] }\left( h,k_a\left( \left[
d_0\right] ,y\right) \right) =i_a\left( h\cdot d_0,y\right) 
\end{equation}
for all arguments.

As $X$ is simply connected, the action $\hat \Phi $ has a global moment map $%
\hat J$ satisfying 
\begin{equation}
\label{pp6t1nonsim3}\frac{\partial \hat \Phi }{\partial G}A\TeXButton{i}
{\intmul}p^{*}\omega +d\left\langle \hat J,A\right\rangle =0\quad . 
\end{equation}
Using the arguments in section \ref{LocalMomentMaps} we find that 
\begin{equation}
\label{pp6t1nonsim4}\frac{\partial \tilde \Phi }{\partial G}A\TeXButton{i}
{\intmul}q^{*}\omega +d\left\langle \hat J\circ j_{a,\left[ d_0\right]
,h},A\right\rangle =0\quad . 
\end{equation}
Hence, introducing the quantities $\tilde J_{a,\left[ d_0\right] ,h}\equiv
\hat J\circ j_{a,\left[ d_0\right] ,h}$, and taking into account that the
trivializations $\left( j_{a,\left[ d_0\right] }\right) $ define an ${\cal H}
$-valued $1$-\u Cech cocycle $\left( \hat g_{a,\left[ d_0\right]
\,;\,b,\left[ d_0^{\prime }\right] }\right) $ by 
\begin{equation}
\label{pp6t1nonsim5}j_{b,\left[ d_0^{\prime }\right] }^{-1}\circ j_{a,\left[
d_0\right] }\left( h,z\right) =\left( h\cdot \hat g_{a,\left[ d_0\right]
\,;\,b,\left[ d_0^{\prime }\right] },z\right) \quad , 
\end{equation}
we see that the collection $\left( \tilde J_{a,\left[ d_0\right]
,h}\,;\,\hat g_{a,\left[ d_0\right] \,;\,b,\left[ d_0^{\prime }\right]
}\right) $ defines a local moment map for the action $\tilde \Phi $ with
respect to $q^{*}\omega $. Similarly, the equation 
\begin{equation}
\label{pp6t1nonsim6}\frac{\partial \Phi }{\partial G}A\TeXButton{i}{\intmul}%
\omega +d\left\langle \hat J\circ i_{a,d},A\right\rangle =0 
\end{equation}
shows that the collection $\left( J_{a,d}\,;\,g_{ab}\right) $, where $%
J_{a,d}\equiv \hat J\circ i_{a,d}$, and $\left( g_{ab}\right) $ is the $%
{\cal D}$-valued $1$-\u Cech cocycle with respect to $\left( i_a\right) $,
is a local moment map for the action $\Phi $ with respect to $\omega $. The
relation between these two local moment maps is easily found using (\ref
{pp6t1nonsim2}) to be 
\begin{equation}
\label{pp6t1nonsim7}\tilde J_{a,\left[ d_0\right] ,h}\circ k_{a,\left[
d_0\right] }=J_{a,h\cdot d_0}\quad . 
\end{equation}

\section{$G$-state spaces and moment maps \label{GStateSpaces}}

Finally, in this section we discuss our concept of a $G$-state space. This
is an identification space based on a partitioning of a symplectic manifold
into connected subsets, on each of which a given global moment map is
constant. These connected subsets are then invariant under the Hamiltonian
flow associated with every Hamiltonian $h$ that commutes with the \thinspace 
$G$-action $\hat \Phi $. This construction coincides with the first step in
a Marsden-Weinstein reduction of the symplectic manifold. We first consider
the case where the symplectic manifold is simply connected:

Let $X$ be a simply connected symplectic manifold with symplectic $2$-form $%
\Omega $. Let $\hat \Phi $ be a symplectic action of a Lie group $G$ on $X$.
There exists a global moment map $J$ associated with $\hat \Phi $. To every $%
x\in X$ we now assign the connected component $s\left( x\right) $ of $%
J^{-1}\left( J\left( x\right) \right) $ that contains $x$; i.e., $s\left(
x\right) \subset J^{-1}\left( J\left( x\right) \right) $ is connected (in
the induced topology), and $x\in s\left( x\right) $. The collection of all $%
s\left( x\right) $, as $x$ ranges through $X$, is denoted as $\Sigma _X$.
Then $\Sigma _X$ is an identification space, where $s:X\rightarrow \Sigma _X$
is the identification map. We endow $\Sigma _X$ with the quotient topology
inherited from $X$. One can assume further technical conditions in order to
guarantee that the sets $s\left( x\right) $ are presymplectic manifolds
which give rise to reduced phase spaces; such a reduction is called
Marsden-Weinstein reduction \cite{Woodhouse}. We do not make these
assumptions here, since they are not necessary for our purposes.

By construction, the moment map $J$ is constant on every connected component
of $J^{-1}\left( J\left( x\right) \right) $, and hence descends to the space 
$\Sigma _X$; i.e., there exists a unique map $\iota :\Sigma _X\rightarrow
g^{*}$ satisfying $\iota \circ s=J$. Also, every diffeomorphism $\hat \Phi
_g $ maps connected components of $J^{-1}\left( J\left( x\right) \right) $
into connected components; this follows from formula (\ref{pp6t1equi4}).
Hence, $\hat \Phi $ descends to an action $\hat \phi :G\times \Sigma
_X\rightarrow \Sigma _X$, satisfying 
\begin{equation}
\label{pp6t1GState0}s\circ \hat \Phi _g=\hat \phi _g\circ s\quad , 
\end{equation}
which gives rise to an analogue of formula (\ref{pp6t1equi4}) on $\Sigma _X$%
, 
\begin{equation}
\label{pp6t1GState1}\iota \circ \hat \phi _g=Ad^{*}\left( g\right) \cdot
\iota +\alpha \left( g\right) \quad . 
\end{equation}

In this construction, we have identified all states $x\in X$ which are
mapped into the same value under the moment map, and which can be connected
by a path on which the moment map is constant. $\Sigma _X$ is a $G$-space
with action $\hat \phi $. There is a semi-equivariant map $\iota $ from $%
\Sigma _X$ to the $G$-space $g^{*}$. Furthermore, the connected components $%
s\left( x\right) $ are preserved by any Hamiltonian that commutes with $G$.
This discussion can be summarized in the

\subsection{Theorem and definition}

Let $X$ be a simply connected symplectic manifold, let $\hat \Phi $ be a
symplectic action of a connected Lie group $G$ with Lie algebra $\hat g$ on $%
X$, let $J$ be a global moment map associated with $\hat \Phi $. Then

\begin{description}
\item[(A)]  \quad there exists a space $\Sigma _X$ with a $G$-action $\hat
\phi $, a projection $s:X\rightarrow \Sigma _X$, and a semi-equivariant map $%
\iota :\Sigma _X\rightarrow g^{*}$ satisfying $\iota \circ s=J$ such that (%
\ref{pp6t1GState1}) holds.

\item[(B)]  \quad If $h$ is any Hamiltonian on $X$ satisfying the
Poisson-bracket relations $\left\{ h,\left\langle J,A\right\rangle \right\}
=0$ for all $A\in \hat g$, then the associated Hamiltonian flow $f_t\left(
x\right) $ preserves the sets $s^{-1}\left( \sigma \right) $, $\sigma \in
\Sigma $. In other words, 
\begin{equation}
\label{pp6t1GState2}J\circ f_t\left( x\right) =J\left( x\right) 
\end{equation}
for all $x\in s^{-1}\left( \sigma \right) $ and $t\in \TeXButton{R}
{\mathbb{R}}$.

\item[(C)]  \quad $\Sigma _X$ will be called a $G${\it -state space} for the
pair $\left( X,\hat \Phi \right) $.
\end{description}

\section{The splitting of multiplets \label{SplittingOfMultiplets}}

Now we turn to investigate the relation of the objects defined above to a
similar construction on a non-simply connected manifold $Y$, where $%
p:X\rightarrow Y$ is a universal covering of $Y$, and $p$ is a local
symplectomorphism of symplectic forms $\omega $ on $Y$ and $\Omega \equiv
p^{*}\omega $ on $X$. The first thing to observe is that the diffeomorphisms 
$\gamma $ of the deck transformation group ${\cal D}$ of the covering
descend to $\Sigma _X$: To see this, let $\lambda $ be a path lying entirely
in one of the connected components $s^{-1}\left( \sigma \right) \subset
J^{-1}\left( J\left( x\right) \right) $ [Here we assume that connectedness
implies path-connectedness]. If $\dot \lambda $ denotes its tangent, we have%
$$
\frac d{dt}\left\langle J\circ \gamma \circ \lambda ,A\right\rangle =\dot
\lambda \TeXButton{i}{\intmul}\gamma ^{*}d\left\langle J,A\right\rangle
=-\Omega \left( \frac{\partial \hat \Phi }{\partial G}A,\gamma _{*}\dot
\lambda \right) = 
$$
$$
=-\left( \gamma ^{*}\Omega \right) \left( \gamma _{*}^{-1}\frac{\partial
\hat \Phi }{\partial G}A,\dot \lambda \right) =-\Omega \left( \frac{\partial
\hat \Phi }{\partial G}A,\dot \lambda \right) =\frac d{dt}\left\langle
J\circ \lambda ,A\right\rangle =0\quad . 
$$
Here, the last equation follows from the fact that $\lambda $ lies in a
connected component of $J^{-1}\left( J\left( x\right) \right) $;
furthermore, we have used that $\Omega $ is ${\cal D}$-invariant, and the
action $\hat \Phi $ commutes with ${\cal D}$, since $G$ is connected (this
follows from theorem \ref{ExtensionOfG}). But this result says that $J$ is
constant on the $\gamma $-image of every connected component $J^{-1}\left(
J\left( x\right) \right) $; since this image is connected itself, it must
lie in one of the connected components of $J^{-1}\left( J\circ \gamma \left(
x\right) \right) $. As $\gamma $ is invertible, it follows that $\gamma $
maps connected components onto connected components, and hence descends to a
map $\bar \gamma :\Sigma _X\rightarrow \Sigma _X$ such that 
\begin{equation}
\label{pp6t1GState3}\bar \gamma \circ s=s\circ \gamma \quad . 
\end{equation}
Thus, we have a well-defined action of ${\cal D}$ on $\Sigma _X$. We now
construct a space $\Sigma _Y$ analogous to $\Sigma _X$: Define $\Sigma _Y$
as the quotient $\Sigma _X/{\cal D}$, with projection $q:\Sigma
_X\rightarrow \Sigma _Y$. We note that this is not a covering space in
general, since the action of ${\cal D}$ on $\Sigma _X$ need not necessarily
be free; for example, if $\gamma $ maps one of the connected components $%
s^{-1}\left( \sigma \right) $ onto itself, then $\bar \gamma $ has a
fixpoint on $\Sigma _X$. However, formula (\ref{pp6t1GState3}) implies that
the map $s$ descends to the quotient $\Sigma _Y=\Sigma _X/{\cal D}$, which
means that there exists a unique map $\bar s:Y\rightarrow \Sigma _Y$ such
that 
\begin{equation}
\label{pp6t1GState4}\bar s\circ p=q\circ s\quad . 
\end{equation}
Using (\ref{pp6t1GState3}) and (\ref{pp6t1GState4}) it is easy to see that
the action $\Phi $ of $G$ on $Y$ preserves the $G$-states on $Y$, i.e. the
images $p\circ s^{-1}\left( \sigma \right) =\bar s^{-1}\circ q\left( \sigma
\right) $ of the connected components of $J^{-1}\left( J\left( x\right)
\right) $, where $x\in s^{-1}\left( \sigma \right) $, under $p$: For, let $%
y\in p\circ s^{-1}\left( \sigma \right) $, then there exists an $x\in
s^{-1}\left( \sigma \right) $ with $y=p\left( x\right) $. Then%
$$
\Phi _g\left( y\right) =\Phi _g\circ p\left( x\right) =p\circ \hat \Phi
_g\left( x\right) \in p\circ \hat \Phi _g\circ s^{-1}\left( \sigma \right)
=p\circ s^{-1}\circ \hat \phi _g\left( \sigma \right) =\bar s^{-1}\circ
q\circ \hat \phi _g\left( \sigma \right) \quad , 
$$
which implies that $\bar s\circ \Phi _g\left( y\right) \in q\circ \hat \phi
_g\left( \sigma \right) $ for all $y$ in $\bar s^{-1}\left( q\left( \sigma
\right) \right) $. Hence, $\Phi $ descends to an action $\phi :G\times
\Sigma _Y\rightarrow \Sigma _Y$ with 
\begin{equation}
\label{pp6t1GState5}\phi _g\circ \bar s=\bar s\circ \Phi _g\quad , 
\end{equation}
which is the analogue of (\ref{pp6t1GState0}).

The orbits $\hat \phi _G\left( \sigma \right) $, $\sigma \in \Sigma _X$, and 
$\phi _G\left( \tau \right) $, $\tau \in \Sigma _Y$, are the classical
analogue of carrier spaces of irreducible $G$-representations in the quantum
context \cite{Woodhouse}. However, for every $G$-state $\tau \in \Sigma _Y$
there exists a collection $q^{-1}\left( \tau \right) $ of $G$-states in $%
\Sigma _X$ which are identified under $q$. The elements in the collection $%
q^{-1}\left( \tau \right) $ are labelled by the elements $d$ of the
fundamental group ${\cal D}\simeq \pi _1\left( Y,y\right) $. We call this
phenomenon the ''splitting of (classical) multiplets'' on account of the
multiple-connectedness of the background $Y$. It is basically a consequence
of the fact that the group $G$, when lifted to the covering space $X$, is
extended to a group $\tilde G$ by the deck transformation group ${\cal D}$,
whose group law in the case under consideration is determined by formula (%
\ref{pp6t1lift12}) in theorem \ref{ExtensionOfG}.

\appendix

\section{Appendix}

Here we compile the cohomologies that are used in this work. References are 
\cite{Fulton},\cite{DodsonParker},\cite{Azcarraga}.

\section{$q$-form-valued cohomology on the deck transformation group \label
{FormValuedCohomologyOnD}}

Let $p:X\rightarrow Y$ be a covering of manifolds, where $X$ is simply
connected. The deck transformation group of the covering is ${\cal D}$. Our
analysis is based on formula (\ref{pp6t1form3}), 
\begin{equation}
\label{pp6t1cohomFormVal1}\gamma ^{*}\eta =\eta +d\chi \left( \gamma \right)
\quad , 
\end{equation}
where $\eta $ is the potential for a $q$-form $d\eta $ which is the
pull-back of a closed $q$-form $\omega $ on $Y$, i.e. $d\eta =p^{*}\omega $.
Then $d\eta $ is ${\cal D}$-invariant, as follows from proposition \ref
{DeckInvariant}, but $\eta $ is not, as follows from the last equation. In
particular this means that for $\chi \neq 0$, $\eta $ is not the pull-back
under $p^{*}$ of a form on $Y$. Now (\ref{pp6t1cohomFormVal1}) defines a
cochain in a cohomology on ${\cal D}$ defined as follows (our notation
conventions are those of \cite{Azcarraga}): An $n$-cochain $\alpha _n$ is a
map $\alpha _n:{\cal D}^n\rightarrow \Lambda ^{*}\left( X\right) $, where $%
\Lambda ^{*}\left( X\right) =\bigoplus\limits_{q\ge 0}\Lambda ^q\left(
X\right) $ denotes the ring of differential forms on $X$. The deck
transformation group ${\cal D}$ acts via pull-back of elements $\gamma $ on
forms: ${\cal D}\times \Lambda ^{*}\left( X\right) \ni \left( \gamma ,\alpha
\right) \mapsto \gamma ^{*}\alpha $. This is a {\bf right} action, in
contrast to the the cohomologies to be discussed below. A zero-cochain $%
\alpha _0$ is an element of $\Lambda ^{*}\left( X\right) $. The coboundary
operator $\delta $ in this cohomology is defined to act on $0$-, $1$-, $2$%
-cochains according to 
\begin{equation}
\label{pp6t1co6}
\begin{array}{rcl}
\left( \delta \alpha _0\right) \left( \gamma \right) & = & \gamma ^{*}\alpha
_0-\alpha _0\quad , \\ 
\left( \delta \alpha _1\right) \left( \gamma _1,\gamma _2\right) & = & 
\gamma _2^{*}\alpha _1\left( \gamma _1\right) -\alpha _1\left( \gamma
_1\gamma _2\right) +\alpha _1\left( \gamma _2\right) \quad , \\ 
\left( \delta \alpha _2\right) \left( \gamma _1,\gamma _2,\gamma _3\right) & 
= & \gamma _3^{*}\alpha _2\left( \gamma _1,\gamma _2\right) +\alpha _2\left(
\gamma _1\gamma _2,\gamma _3\right) - \\  
&  & -\alpha _2\left( \gamma _1,\gamma _2\gamma _3\right) -\alpha _2\left(
\gamma _2,\gamma _3\right) \quad , 
\end{array}
\end{equation}
and $\delta $ is nilpotent, $\delta \circ \delta =0$, as usual. Now consider
the $0$-cochain $\eta $ in equation (\ref{pp6t1cohomFormVal1}). With the
help of (\ref{pp6t1co6}), equation (\ref{pp6t1cohomFormVal1}) can be
expressed as 
\begin{equation}
\label{pp6t1co7}\delta \eta =d\chi \quad , 
\end{equation}
where it is understood that $\chi $ is a function of arguments in the set $%
{\cal D}\times X$. The commutativity $\left[ \delta ,d\right] =0$ and the
nilpotency $\delta ^2=0$ and $d^2=0$ of the coboundary operators give rise
to a chain of equations similar to (\ref{pp6t1co7}): Applying $\delta $ to (%
\ref{pp6t1co7}) gives 
\begin{equation}
\label{pp6t1co8}d\left( \delta \chi \right) =0\quad . 
\end{equation}
Since $X$ is simply connected 
\begin{equation}
\label{pp6t1co9}\delta \chi =d\chi ^{\prime } 
\end{equation}
for some $\chi ^{\prime }$. Now the process can be repeated with the last
equation, etc.

\section{\u Cech cohomology \label{CechCohomology}}

The definitions in this section are based on \cite{Fulton},\cite
{DodsonParker}.

Let $Y$ be a topological space and ${\cal D}$ be a discrete group. Let $%
{\cal V}\equiv \left\{ V_a\mid a\in A\right\} $ be an open cover of $Y$ such
that every $V_a$ is admissible. [If $Y$ is a manifold, we can assume that $A$
is countable, and every $V_a$ is simply connected.] A function $%
g:Y\rightarrow {\cal D}$ on the topological space $Y$ is called {\it locally
constant} if every point $y\in Y$ possesses a neighbourhood $V$ on which the
restriction of $g$ is constant.

Let ${\cal S}^0$ denote the sum ${\cal S}^0\equiv \bigsqcup\limits_{a\in
A}V_a$; let ${\cal S}^1$ denote the sum ${\cal S}^1\equiv
\bigsqcup\limits_{a,b}V_a\cap V_b$, for all $a,b$ for which $V_a\cap V_b\neq
\emptyset $, allowing for $a=b$. We denote the images of $V_a$, $V_a\cap V_b$%
, ..., under the associated injections simply by $\left( a\right) $, $\left(
a,b\right) $, etc. [We recall that the set underlying a sum $\Sigma =B\sqcup
C$ is the disjoint union of $B$ and $C$. Furthermore, if $i:B\rightarrow
\Sigma $, $j:C\rightarrow \Sigma $ are the injections, then $i\left(
B\right) $ and $j\left( C\right) $ are both open and closed in $\Sigma $,
which means that $\Sigma $ is disconnected. Hence if each $V_a$ is connected
in $Y$, then a locally constant function on ${\cal S}^0$ is constant on the
images of all $V_a$ under the appropriate injection; similarly, a locally
constant function on ${\cal S}^1$ is constant on the images of $V_a\cap V_b$
under injection. Therefore in this case, a locally constant function on $%
{\cal S}^0,{\cal S}^1$ is constant on all $\left( a\right) $, and $\left(
a,b\right) $, respectively]. A locally constant function $f_0:{\cal S}%
^0\rightarrow {\cal D}$ is called a $0$-\u Cech-cochain (with respect to $%
{\cal V}$). A locally constant function $f_1:{\cal S}^1\rightarrow {\cal D}$
is called a $1$-\u Cech-cochain (with respect to ${\cal V}$). A $1$-\u
Cech-cochain $f_1$ is called a $1${\it -\u Cech-cocycle} if

\begin{description}
\item[(Coc1)]  \ $\left. f_1\right| \left( a,a\right) =e$, where $e$ is the
identity in ${\cal D}$,

\item[(Coc2)]  \ $\left. f_1\right| \left( b,a\right) =\left.
f_1^{-1}\right| \left( a,b\right) $, $f_1^{-1}$ denoting the inverse of $f_1$
in ${\cal D}$;
\end{description}

and for all $V_a,V_b,V_c$ for which $V_a\cap V_b\cap V_c\neq \emptyset $ it
is true that

\begin{description}
\item[(Coc3)]  \ $\left. f_1\right| \left( a,c\right) =\left[ \left.
f_1\right| \left( a,b\right) \right] \cdot \left[ \left. f_1\right| \left(
b,c\right) \right] $,
\end{description}

where $\left( a,b\right) $, $\left( b,a\right) $, $\left( a,c\right) $,
etc., are to be regarded as disjoint subsets of ${\cal S}^1$.

Two $1$-\u Cech-cocycles $f,f^{\prime }$ are said to be {\it cohomologous}
if there exists a $0$-\u Cech-cochain $h$ such that

\begin{description}
\item[(Coh)]  \ $\left. f^{\prime }\right| \left( a,b\right) =\left[ \left.
h^{-1}\right| \left( a\right) \right] \cdot \left[ \left. f\right| \left(
a,b\right) \right] \cdot \left[ \left. h\right| \left( b\right) \right] $.
\end{description}

The property of being cohomologous defines an equivalence relation amongst
all $1$-\u Cech-cocycles with respect to ${\cal V}$; the equivalence classes
are called {\it first \u Cech cohomology classes on }${\cal V}$ with
coefficients in ${\cal D}$. The set of these classes is denoted as $%
H^1\left( {\cal V};{\cal D}\right) $.

For $n>1$, the $n$-th cohomology class is described more readily when ${\cal %
D}$ is Abelian. Assuming this, an $n$-{\it \u Cech-cochain} $f_n$ with
coefficients in ${\cal D}$ is a locally constant map $f_n:{\cal S}%
^n\rightarrow {\cal D}$, where ${\cal S}^n$ is the topological sum ${\cal S}%
^n\equiv \bigsqcup\limits_{a_0,\ldots ,a_n}V_{a_0}\cap V_{a_1}\cdots \cap
V_{a_n}$, for all $a_0,\ldots ,a_n$ for which $V_{a_0}\cap V_{a_1}\cdots
\cap V_{a_n}\neq \emptyset $. The coboundary operator $\delta $ sends $n$%
-cochains $f_n$ to $\left( n+1\right) $-cochains $\delta f_n$ defined by 
\cite{DodsonParker} 
\begin{equation}
\label{pp6t1cechcohom1}\left. \delta f_n\right| \left( a_0,\ldots
,a_{n+1}\right) =\sum_{k=0}^{n+1}\left( -1\right) ^k\cdot \left. f_n\right|
\left( a_0,\ldots ,\widehat{a_k},\ldots ,a_{n+1}\right) \quad , 
\end{equation}
where $\widehat{a_k}$ means that this argument has to be omitted, and $-f_n$
denotes the inverse of $f_n$ in the Abelian group ${\cal D}$. $\delta $ is
nilpotent, $\delta _{n+1}\circ \delta _n=0$. As usual, $n$-cocycles are
elements in $\ker \delta _n$, $n$-coboundaries are elements in $\left. {\rm %
im\,}\delta _{n-1}\right. $, and the $n$-th \u Cech cohomology group on $%
{\cal V}$ is the quotient $H^n\left( {\cal V};{\cal D}\right) =\ker \delta
_n/\left. {\rm im\,}\delta _{n-1}\right. $. Obviously, statements
(Coc1-Coc3) and (Coh) above generalize this pattern to non-Abelian groups $%
{\cal D}$.

\section{${\cal D}$-coverings \label{DCoverings}}

This and the next section are mainly based on \cite{Fulton}.

Let ${\cal D}$ denote a discrete group. Let $X$ be a topological space on
which ${\cal D}$ acts properly discontinuously and freely. Let $%
p:X\rightarrow X/{\cal D}$ denote the projection onto the space of orbits,
endowed with the quotient (final) topology. Then $p$ is a covering map.

Generally, if a covering $p:X\rightarrow Y$ arises in this way from a
properly discontinuous and free action ${\cal D}$ on $X$, we call the
covering a ${\cal D}${\it -covering}. Given two coverings $p:X\rightarrow Y$
and $p^{\prime }:X^{\prime }\rightarrow Y^{\prime }$ of topological spaces,
a homeomorphism $\phi :X\rightarrow X^{\prime }$ is called {\it isomorphism
of coverings} if $\phi $ is fibre-preserving, $p^{\prime }\circ \phi =p$. An 
{\it isomorphism of ${\cal D}$-coverings} is an isomorphism of coverings
that commutes with the actions of ${\cal D}$, i.e. $\phi \left( d\cdot
x\right) =d\cdot \phi \left( x\right) $; in this case we also say that $\phi 
$ is ${\cal D}${\it -equivariant}, and we say that the spaces involved are $%
{\cal D}${\it -isomorphic}. The {\it trivial} ${\cal D}${\it -covering} of $%
Y $ is the Cartesian product ${\cal D}\times Y$ together with projection
onto the second factor as covering map, and ${\cal D}$ acts on $\left(
d,y\right) \in {\cal D}\times Y$ by left multiplication on the first factor, 
$\left( d^{\prime },\left( d,y\right) \right) \mapsto \left( d^{\prime
}d,y\right) $. An isomorphism $i:{\cal D}\times Y\rightarrow X$ of the
trivial ${\cal D}$-covering onto a ${\cal D}$-covering $X$ is called a {\it %
trivialization of }$X$.

Let $p:X\rightarrow X/{\cal D}=Y$ be a ${\cal D}$-covering. Let $V\subset Y$
be an admissible connected open set in $Y$. A choice of a connected
component $U\subset p^{-1}\left( V\right) $ defines a {\it trivialization} $%
i:{\cal D}\times V\rightarrow p^{-1}\left( V\right) $ of the ${\cal D}$%
-covering $p:p^{-1}\left( V\right) \rightarrow V$ as follows: For $\left(
d,y\right) \in {\cal D}\times V$, let $i\left( d,y\right) \equiv d\cdot
\left( \left. p\right| U\right) ^{-1}\left( y\right) $. Then $i$ is
evidently a bijection and hence a homeomorphism; furthermore it is
fibre-preserving, since projection onto the second factor of $\left(
d,y\right) $ yields the same as $p\circ i$ applied to $\left( d,y\right) $;
and it is ${\cal D}$-equivariant by definition. This says that a ${\cal D}$%
-covering is trivial over each admissible neighbourhood $V\subset Y$. A
different choice $U^{\prime }$ of connected components in $p^{-1}\left(
V\right) $ defines a trivialization $i^{\prime }:{\cal D}\times V\rightarrow
p^{-1}\left( V\right) $ with $i^{\prime -1}\circ i\left( d,y\right) =\left(
d\cdot g\left( y\right) ,y\right) $, where $g:V\rightarrow {\cal D}$ is
continuous, and hence constant on every connected subset of $V$, since $%
{\cal D}$ is discrete.

\section{\u Cech cohomology and the glueing of ${\cal D}$-coverings \label
{CechAndGlueing}}

Let $p:X\rightarrow X/{\cal D}=Y$ be a ${\cal D}$-covering; let ${\cal V}$
be an open cover of $Y$ by admissible subsets $V\subset Y$. For sufficiently
simple spaces such as manifolds it can be assumed that every $V\in {\cal V}$
is simply connected in $Y$ and path-connected. As explained in the last
section, a choice of connected component $U_a\subset p^{-1}\left( V_a\right) 
$ in the inverse image of every $V_a$, $a\in A$, gives rise to a set of
local trivializations $i_a:{\cal D}\times V_a\rightarrow p^{-1}\left(
V_a\right) $, which, in turn, define a collection $\left( g_{ab}\right) $ of
transition functions $g_{ab}:V_a\cap V_b\rightarrow {\cal D}$, $%
i_b^{-1}\circ i_a\left( d,y\right) =\left( d\cdot g_{ab}\left( y\right)
,y\right) $. If all $V_a$ are connected, the transition functions $g_{ab}$
are constant due to continuity. It is easily seen that the $\left(
g_{ab}\right) $ satisfy

\begin{description}
\item[(Trans1)]  \ $g_{aa}=e$,

\item[(Trans2)]  \ $g_{ba}=g_{ab}^{-1}$,

\item[(Trans3)]  \ $g_{ac}=g_{ab}\cdot g_{bc}$,
\end{description}

the last equation following from $i_a^{-1}i_ci_c^{-1}i_bi_b^{-1}i_a=id$,
whenever $V_a\cap V_b\cap V_c\neq \emptyset $. Comparison with (Coc1-Coc3)
in section \ref{CechCohomology} shows that the collection $\left(
g_{ab}\right) $ defines a $1$-\u Cech-cocycle on ${\cal V}$. Now suppose we
choose different trivializations $i_a^{\prime }$. Then the trivializations
are related by $i_a^{\prime -1}\circ i_a\left( d,y\right) =\left( d\cdot
h_a\left( y\right) ,y\right) $, with a collection $\left( h_a\right) $ of
locally constant functions $h_a:V_a\rightarrow {\cal D}$, which defines a $0$%
-\u Cech-cochain on ${\cal V}$, as explained in section \ref{CechCohomology}%
. The transition functions $\left( g_{ab}^{\prime }\right) $ associated with 
$\left( i_a^{\prime }\right) $ are defined by $i_b^{\prime -1}\circ
i_a^{\prime }\left( d,y\right) =\left( d\cdot g_{ab}^{\prime },y\right) $;
on the other hand, from the definition of $\left( h_a\right) $, we find that 
$i_a^{\prime }\left( d,y\right) =i_a\left( d\cdot h_a^{-1},y\right) $, which
implies that $i_b^{\prime -1}\circ i_a^{\prime }\left( d,y\right) =\left(
d\cdot h_a^{-1}g_{ab}h_b,y\right) $. Thus, 
\begin{equation}
\label{pp6t1appCechCoh}g_{ab}^{\prime }=h_a^{-1}\cdot g_{ab}\cdot h_b\quad . 
\end{equation}
Statement (Coh) in section \ref{CechCohomology} now shows that the cocycles $%
\left( g_{ab}\right) $ and $\left( g_{ab}^{\prime }\right) $ are
cohomologous. This means that a ${\cal D}$-covering $p:X\rightarrow X/{\cal D%
}=Y$ determines a unique \u Cech cohomology class in $H^1\left( {\cal V;D}%
\right) $. -- Furthermore, let $q:Z\rightarrow Z/{\cal D}=Y$ be another $%
{\cal D}$-covering of $Y$ such that there exists an isomorphism $\phi
:X\rightarrow Z$ of ${\cal D}$-coverings. Assume that $i_a:{\cal D}\times
V_a\rightarrow p^{-1}\left( V_a\right) \subset X$ is the trivialization over 
$V_a$ in the ${\cal D}$-covering $p:X\rightarrow Y$. Then $\left( d,y\right)
\mapsto \phi \circ i_a\left( d,y\right) $ is a trivialization over $V_a$ in
the covering $q:Z\rightarrow Y$ with transition functions $g_{ab}^{\prime
}=g_{ab}$. Any other trivialization of $q:Z\rightarrow Y$ gives a
cohomologous cocycle. Thus, we have found that ${\cal D}$-coverings which
are ${\cal D}$-isomorphic define a unique \u Cech cohomology class in $%
H^1\left( {\cal V;D}\right) $.

-- Conversely, we want to show that a cohomology class represented by $%
\left( g_{ab}\right) $ defines a ${\cal D}$-covering of $Y$ up to ${\cal D}$%
-isomorphisms. Let $p:X\rightarrow Y=X/{\cal D}$, $p^{\prime }:X^{\prime
}\rightarrow Y=X^{\prime }/{\cal D}$ be two ${\cal D}$-coverings with
trivializations $i_a:{\cal D}\times V_a\rightarrow p^{-1}\left( V_a\right) $%
, $i_a^{\prime }:{\cal D}\times V_a\rightarrow p^{\prime -1}\left(
V_a\right) $ and associated cocycles $\left( g_{ab}\right) $, $\left(
g_{ab}^{\prime }\right) $ defined by $i_b^{-1}\circ i_a\left( d,y\right)
=\left( d\cdot g_{ab},y\right) $, $i_b^{\prime -1}\circ i_a^{\prime }\left(
d,y\right) =\left( d\cdot g_{ab}^{\prime },y\right) $, such that $%
g_{ab}^{\prime }=h_a^{-1}\cdot g_{ab}\cdot h_b$, where $\left(
h_a:V_a\rightarrow {\cal D}\right) $ is a $0$-\u Cech-cochain on ${\cal V}$.
It follows that $i_a^{\prime }\left( d\cdot h_a,y\right) =i_b^{\prime
}\left( d\cdot g_{ab}\cdot h_b,y\right) $. On the sets ${\cal D}\times V_a$
we now define a collection of functions $k_a:{\cal D}\times V_a\rightarrow
X^{\prime }$ determined by $k_a\left( d,y\right) =i_a^{\prime }\left( d\cdot
h_a,y\right) $. By construction, all $k_a$ are continuous. Furthermore, if $%
\left( d,y\right) \in {\cal D}\times V_a$ and $\left( d^{\prime },y^{\prime
}\right) \in {\cal D}\times V_b$ are identified under the trivializations $%
i_a$, $i_b$, so that $i_a\left( d,y\right) =i_b\left( d^{\prime },y^{\prime
}\right) $, then $k_a\left( d,y\right) $ and $k_b\left( d^{\prime
},y^{\prime }\right) $ coincide; for, in this case, we must have $\left(
d^{\prime },y^{\prime }\right) =\left( d\cdot g_{ab},y\right) $, and hence%
$$
k_b\left( d^{\prime },y^{\prime }\right) =i_b^{\prime }\left( d^{\prime
}\cdot h_b,y^{\prime }\right) =i_b^{\prime }\left( d\cdot g_{ab}\cdot
h_b,y\right) =i_b^{\prime }\left( dh_a\cdot g_{ab}^{\prime },y\right) = 
$$
$$
=i_a^{\prime }\left( d\cdot h_a,y\right) =k_a\left( d,y\right) \quad . 
$$
Now the set of trivializations $\left( i_a:{\cal D}\times V_a\rightarrow
X\right) $ as defined here can be regarded as a collection of identification
maps $\left( i_a\right) $ on the sets ${\cal D}\times V_a$, where ${\cal D}$
has the discrete topology, and the $V_a$ have the topology induced from $Y$.
The identification space $X$ has the $\phi $-universal property \cite{Brown}
that for every topological space $X^{\prime }$ and any collection of
continuous functions $\left( k_a:{\cal D}\times V_a\rightarrow X^{\prime
}\right) $ which coincide on elements $\left( d,y\right) $, $\left(
d^{\prime },y^{\prime }\right) $ which are identified in the identification
space $\left( i_a:{\cal D}\times V_a\rightarrow X\right) $, there exists a
unique continuous map $\psi :X\rightarrow X^{\prime }$ such that $\psi \circ
i_a=k_a$. Since the arguments leading to this result can be reversed, it
follows that $\psi $ has a continuous inverse, and hence is a homeomorphism.
Locally, we have $i_a^{\prime -1}\circ \psi \circ i_a\left( d,y\right)
=\left( d\cdot h_a,y\right) $, which says that $\psi $ is fibre-preserving,
and hence is a covering isomorphism. The same formula shows that $\psi $ is $%
{\cal D}$-equivariant. Altogether, therefore, $\psi $ is an isomorphism of $%
{\cal D}$-coverings.

The last two paragraphs therefore prove that there is a 1--1 correspondence
between ${\cal D}$-coverings of $Y$ which are ${\cal D}$-isomorphic, and
cohomology classes in the first \u Cech cohomology group $H^1\left( {\cal V;D%
}\right) $ on ${\cal V}$ with values in ${\cal D}$.

\section{Deck-transformation-valued cohomology on $G$ \label
{DeckTransformationValuedCohomOnG}}

Consider the scenario of theorem \ref{ExtensionOfG}; there it was shown that
when the action $\phi :G\times Y\rightarrow Y$ of the Lie group $G$ is
lifted to a smooth map $\hat \phi :G\times X\rightarrow X$ satisfying
(L1,L2), then the set $\left\{ \hat \phi _g\mid g\in G\right\} $ usually no
longer closes into a group, but is extended to a larger group $\tilde G$
which contains ${\cal D}$ as a normal subgroup such that $\tilde G/{\cal D}%
=G $. The deviation from closure was measured by the map 
\begin{equation}
\label{pp6t1co10}\Gamma :G\times G\rightarrow {\cal D}\quad ,\quad \left(
g,h\right) \mapsto \Gamma \left( g,h\right) \equiv \hat \phi _g\hat \phi _h%
\widehat{\phi _{gh}}^{-1}\quad , 
\end{equation}
see (\ref{pp6t1lift8}). Furthermore, we have a map 
\begin{equation}
\label{pp6t1co11}b\left( g\right) :{\cal D}\rightarrow {\cal D}\quad ,\quad
\gamma \mapsto b\left( g\right) \gamma \equiv \hat \phi _g\circ \gamma \circ
\hat \phi _g^{-1}\quad . 
\end{equation}
In the discussion following formula (\ref{pp6t1lift7}) it was pointed out
that $b:G\rightarrow Aut\left( {\cal D}\right) $ usually is not a
representation; here we show that if ${\cal D}$ is Abelian, then $b$ {\bf is}
a representation, and hence defines a left action 
\begin{equation}
\label{pp6t1co12}G\times {\cal D}\rightarrow {\cal D\quad },\quad \left(
g,\gamma \right) \mapsto b\left( g\right) \gamma \quad , 
\end{equation}
of $G$ on ${\cal D}$. To see this, consider the expression%
$$
b\left( gg^{\prime }\right) \gamma =\hat \phi _{gg^{\prime }}\gamma \hat
\phi _{gg^{\prime }}^{-1}\quad ; 
$$
using (\ref{pp6t1co10},\ref{pp6t1co11}) this becomes%
$$
\hat \phi _{gg^{\prime }}\gamma \hat \phi _{gg^{\prime }}^{-1}=\Gamma
^{-1}\left( g,g^{\prime }\right) \left[ b\left( g\right) \circ b\left(
g^{\prime }\right) \gamma \right] \Gamma \left( g,g^{\prime }\right) \quad ; 
$$
but the expression $b\left( g\right) \circ b\left( g^{\prime }\right) \gamma 
$ in square brackets is an element of ${\cal D}$, as are the $\Gamma $'s.
Hence, since ${\cal D}$ is Abelian, the last expression is%
$$
\Gamma ^{-1}\left( g,g^{\prime }\right) \left[ b\left( g\right) \circ
b\left( g^{\prime }\right) \gamma \right] \Gamma \left( g,g^{\prime }\right)
=b\left( g\right) \circ b\left( g^{\prime }\right) \gamma \quad , 
$$
which proves 
\begin{equation}
\label{pp6t1co13}b\left( gg^{\prime }\right) =b\left( g\right) \circ b\left(
g^{\prime }\right) \quad . 
\end{equation}

In the sequel we use an additive notation for the group law in ${\cal D}$;
i.e. $\left( \gamma ,\gamma ^{\prime }\right) \mapsto \gamma +\gamma
^{\prime }\in {\cal D}$. We now introduce a ${\cal D}$-valued cohomology on $%
G$ as follows: $n$-cochains $\alpha _n$ are maps $G^n\rightarrow {\cal D}$; $%
0$-cochains are elements of ${\cal D}$. The coboundary operator $\delta $ is
defined to act on $0$-, $1$-, $2$-cochains according to 
\begin{equation}
\label{pp6t1co14}
\begin{array}{rcl}
\left( \delta \alpha _0\right) \left( g\right) & = & b\left( g\right) \alpha
_0-\alpha _0\quad , \\ 
\left( \delta \alpha _1\right) \left( g,h\right) & = & b\left( g\right)
\alpha _1\left( h\right) -\alpha _1\left( gh\right) +\alpha _1\left(
g\right) \quad , \\ 
\left( \delta \alpha _2\right) \left( g,h,k\right) & = & b\left( g\right)
\alpha _2\left( h,k\right) +\alpha _2\left( g,hk\right) - \\  
&  & -\alpha _2\left( gh,k\right) -\alpha _2\left( g,h\right) \quad . 
\end{array}
\end{equation}
This is well-defined, since (\ref{pp6t1co13}) says that $b$ is now an
action. We denote the sets of $n$-cochains, -cocycles, -coboundaries, and $n$%
-cohomology groups by $C^n\left( G,{\cal D}\right) $, $Z^n\left( G,{\cal D}%
\right) $, $B^n\left( G,{\cal D}\right) $, and $H^n\left( G,{\cal D}\right)
=Z^n\left( G,{\cal D}\right) /B^n\left( G,{\cal D}\right) $.

\TeXButton{Abst}{\vspace{0.8ex}}--- Another cohomology on $G$ and $\hat g$
that occurs in studying moment maps is the

\section{$g^{*}$-valued cohomology on $G$ \label{gStarOnGroup}}

$n${\it -}cochains{\it \ }are smooth maps $\alpha _n:G^n\rightarrow g^{*}$. $%
G$ acts on $g^{*}$ via the coadjoint representation $Ad^{*}$ of $G$ on $%
g^{*} $; this is a left action, see the beginning of the appendix. The set
of all $g^{*}$-valued $n$-cochains is denoted by $C^n\left( G,g^{*}\right) $%
. The coboundary operator $\delta :C^n\rightarrow C^{n+1}$ acts on $0$-, $1$%
-, $2$-cochains $\alpha _0$, $\alpha _1$, $\alpha _2$ according to 
\begin{equation}
\label{pp6t1co4}
\begin{array}{rcl}
\left( \delta \alpha _0\right) \left( g\right) & = & Ad^{*}\left( g\right)
\alpha _0-\alpha _0\quad , \\ 
\left( \delta \alpha _1\right) \left( g,h\right) & = & Ad^{*}\left( g\right)
\alpha _1\left( h\right) -\alpha _1\left( gh\right) +\alpha _1\left(
g\right) \quad , \\ 
\left( \delta \alpha _2\right) \left( g,h,k\right) & = & Ad^{*}\left(
g\right) \alpha _2\left( h,k\right) +\alpha _2\left( g,hk\right) - \\  
&  & -\alpha _2\left( gh,k\right) -\alpha _2\left( g,h\right) \quad , 
\end{array}
\end{equation}
etc.. The set of all $n$-cocycles is denoted by $Z^n\left( G,g^{*}\right) $,
the set of all $n$-coboundaries is denoted as $B^n\left( G,g^{*}\right) $.
The $n$-th cohomology group of $G$ with values in $g^{*}$ is the quotient $%
H^n\left( G,g^{*}\right) =Z^n\left( G,g^{*}\right) /B^n\left( G,g^{*}\right) 
$.


\begin{thebibliography}{99}
\bibitem{AzTown}  de Azcarraga, J.A.; Gauntlett, J.P.; Izquierdo, J.M., and
Townsend, P.K. (1989): ''Topological extensions of the supersymmetry algebra
for extended objects''. Phys. Rev. Lett. {\bf 63}, 2443--2446.

\bibitem{Woodhouse}  N.M.J. Woodhouse, ''Geometric Quantization''. Oxford
Science Publications, 1991.

\bibitem{Fulton}  William Fulton, ''Algebraic Topology''. Springer Verlag,
1995.

\bibitem{Jaehnich}  Klaus J\"ahnich, ''Topology''. Springer-Verlag, 1980.

\bibitem{ONeill}  Barrett O'Neill, ''Semi-Riemannian Geometry with
Applications to Relativity''. Academic Press, 1983.

\bibitem{Wolf}  J.A. Wolf, ''Spaces of Constant Curvature''. McGraw Hill,
1967.

\bibitem{Brown}  Ronald Brown, ''Topology''. Ellis Harwood Limited, 1988.

\bibitem{GuillStern}  Victor Guillemin, Shlomo Sternberg, ''Symplectic
techniques in physics''. Cambridge University Press, 1984.

\bibitem{AbrahamMarsden}  Ralph Abraham, Jerrold E. Marsden. ''Foundations
of Mechanics''. The Benjamin/Cummings Publishing Company, Inc., 1978.

\bibitem{CramPir}  M. Crampin and F.A.E. Pirani, ''Applicable Differential
Geometry''. Cambridge University Press, 1986.

\bibitem{DodsonParker}  C.T.J. Dodson and Phillip E. Parker, ''A User's
Guide to Algebraic Topology''. Kluwer Academic Publishers.

\bibitem{Azcarraga}  Jose de Azcarraga and Jose M. Izquierdo, ''Lie groups,
Lie algebras, cohomology and some applications in physics''. Cambridge
University Press, 1995.
\end{thebibliography}
\end{document}